\documentclass[12pt]{article}
\usepackage[dvips]{color}
\usepackage{epsfig}
\usepackage{amsmath}
\usepackage{graphicx}

\begin{document}
\title{\bf Phase transition of modified Horndeski gravity with new method}
\author{{\small{M. Rostami$^{a,}$\thanks{Email: M.rostami@iau-tnb.ac.ir}, \hspace{1mm} J. Sadeghi$^{b,}$\thanks{Email: pouriya@ipm.ir}, \hspace{1mm} S. Miraboutalebi$^{a,}$\thanks{Email: S-mirabotalebi@iau-tnb.ac.ir},\hspace{1mm} B. Pourhassan$^{c,}$\thanks{Email: b.pourhassan@du.ac.ir} \hspace{1mm} and \hspace{1mm} A. A. Masoudi$^{d,}$\thanks{Email: masoudi@alzahra.ac.ir}}}\\
$^{a}${\small {\em  Department of Physics, North Tehran Branch, Islamic Azad University, Tehran, Iran.}}\\
$^{b}${\small {\em Department of Physics, Faculty of Basic Sciences, University of Mazandaran, Babolsar, Iran.}}\\
$^{c}${\small {\em  School of Physics, Damghan University, P. O. Box 3671641167, Damghan, Iran.}}\\
$^{d}${\small {\em Department of Physics, Alzahra University, Tehran, Iran.}}} \maketitle

\begin{abstract}
\noindent In this paper, we investigate the critical points of the $ P-V $ diagram and the phase transitions of the Horndenski black holes. In fact, the usual Horndeski black holes do not have $ P-V $ critical points, hence do not show any phase transitions. However, we successfully modify the Horndeski black hole solution to obtain such a phase transition behavior. This modified black holes solution is satisfied by the equation of state of liquid-gas phase transition. Also, we study the thermodynamics of our modified Horndeski black hole by applying a new method based on the equation of state that originated from the slope of temperature versus entropy. This new prescription provides us a simple powerful way to study the critical behavior and the phase transition of the black holes and concludes the novel results. The analytical interpretation of possible phase transition points leads us to set some nonphysical range on the horizon radius for the black hole.
\end{abstract}

\newpage
\section{Introduction}
The general theory of relativity can describe very well the classical phenomena of the black holes.  However, the quantum aspects of the black holes can be defined and studied in the framework of the quantum gravity. These aspects play an important role to understand quantum gravity and improve our knowledge about it. However, in order to ascribe quantum aspects to the classical black holes, we need to apply some modifications to the ordinary black hole solutions. In this regard, we consider the Horndeski black holes \cite{1}, and specifically we are interested in the thermodynamics of AdS space-time \cite{2, 3}.\\
Perhaps, Hawking and Bekenstein were the first persons to find and establish beautiful relations between the rules of black hole mechanics and the ordinary laws of thermodynamics. They studied the thermodynamics of black holes and found out that there are relations between the surface gravity, mass, and area of a black hole, on one hand, and its temperature, energy, and entropy respectively, on the other hand \cite{4,5,6,7,8,9,10,11}.\\
The idea of including the variation of cosmological constant $ \Lambda $  in the first law of black hole thermodynamics has been attended recently by several authors and applied to some black holes.
In this case, the parameter $ \Lambda $  is related to pressure and its conjugate variable  namely, volume, while this is a special characteristic to obtain the black hole space-time.
Here, we use the units $ G_{N}=\hbar = c = \kappa = 1$ and identify the pressure with the following expression,
\begin{equation}\label{1}
P = - \frac{\Lambda}{8 \pi} = \frac{3}{8 \pi\ell^{2}},
\end{equation}
where $ \ell$ is the length of AdS space. Also, the corresponding volume is,
\begin{equation}\label{2}
V = ( \frac{\partial M}{\partial P})_{S,Q,J},
\end{equation}
where $ M $ is the black hole mass. Eq.(\ref{1}) indicates that the black hole thermodynamics will be interesting with the presence of a negative cosmological constant that is a specific characteristic of the AdS space, and specifically is so fruitful in holography and  AdS/CFT  correspondence.\\
The critical point in charged AdS black holes shows that such a black hole has a van der Waals fluid behavior \cite{12}. It has been found that the van der Waals fluid is the holographic dual of RN AdS black hole \cite{13}. So, by using the holographic principles, one can study AdS black holes via a van der Waals fluid and discuss some $P - V$ criticality \cite{12}. Also it is found that spinning Kerr-AdS black hole in five dimensions, behave as van der Waals fluid \cite{NPB1}.\\
The van der Waals phase transition and $P - V$ criticality of AdS black holes in general framework are already discussed by the Refs. \cite{Majhi1, Majhi2}, which is extended to the massive gravity by the Ref. \cite{EPJC1} and found that presence of logarithmic correction is necessary to have a holographically dual of van der Waals fluid \cite{PRD1, PLB1}.\\
The equation of state of van der Waals fluid is a popular closed form modification of the ideal gas law. It approximates the behavior of real fluids by taking into account the nonzero size of molecules and the attraction between them.  It is often used to describe the qualitative features of the Liquid-gas phase transition. In that case, the equation reads,
\begin{equation} \label{3}
\left ( P+ \frac{a}{v^2}\right )(v-b) = kT,
\end{equation}
where $ v = \frac{V}{N} $ is the specific volume of the fluid  and $k$ is the Boltzmann
constant. The constant $ b > 0 $ takes into account the nonzero size
of the molecules of a given fluid, whereas the constant $ a > 0 $ ensures  the attraction between them. One can
expand this equation to write it as a cubic equation for $v$,
\begin{equation}\label{4}
Pv^3 - \left ( kT + bP \right ) v^2 + av - ab = 0.
\end{equation}
In order to investigate the $ P - V $  critical points of gas, we need to apply the following condition,
\begin{equation}\label{5}
\left (\frac{\partial P}{\partial v}\right )_{S,Q,J} = 0, \qquad  \left (\frac{\partial^2
P}{\partial v^2 }\right)_{S,Q,J} = 0,
\end{equation}
In the presence of a negative cosmological constant, the asymptotically AdS black hole admit a gauge duality description with dual thermal field theory. Such theory leads us to an interesting phenomenon which is called Hawking and Page phase transition \cite{13,14,15,16,17}. This article is devoted to the phase transitions take place near the critical point.\\
There are different approaches to investigate the phase transition which some of them have studied the behavior of the heat capacity in different ensembles.   Here, we use two major approaches to examine the phase transition.
In the first approach, the changes of the signature in the heat capacity are representing the phase transitions and hence the roots of heat capacity have decisive roles. In the second approach, the divergences of the heat capacity are indicating the phase transitions so that the singular points of the heat capacity become more important \cite{18,19}.\\
The heat capacity is an interesting thermodynamical quantity to determine the stability and instability of the black hole. In general, black hole heat capacity is always negative which shows that the black hole is unstable and have Hawking radiation. But with the presence of charge and rotation parameters of the black hole, the heat capacity can change the sign, and become positive, thus the phase transition occurs. In this article, we use another new method to study the phase transitions in which the critical behavior of the van der Waals gas are obtained by using the slope of $T$ versus $S$ \cite{22, 23, 24}.\\
According to the standard methods,  in the usual extended phase transition space,  one should calculate firstly  $ T = \frac{\partial M}{\partial S} $ to obtain the equation of state. The other calculations then take place by using the state equation. Instead, here applying the new method, we use $ \frac{\partial T }{\partial S } = \frac{\partial^{2} M}{\partial S^{2}} $  to find the equation of state. Then,  by knowing the equation of state,  the thermodynamical quantities of our physical system can be studied.\\
The organization of this paper is as follows: In section 2, we study the standard solution of the Hordeski black hole.  This usual solution does not show critical behavior in its $P-V$ diagram. In section 3, we then generalize this solution to modify this behavior by applying some ansatz.  In section 4, we use the standard method to study the critical points of the $P-V$ diagram for our modified Hordeski black hole and obtain its phase transitions and then employ the thermodynamical relations. In section 5, we apply the new method to obtain the equation of state based on the slope of temperature versus entropy, and calculate the phase transition parameters. We also compare the obtained results with the consequences of the standard method. Also, we investigate the stability and instability of our modified black hole.   Finally, we summarize our results with concluding remarks in the last section.

\section{Horndeski black hole solution}
We begin with the following action \cite{1},
\begin{equation}\label{6}
S =\int{d^4 x \sqrt{-g}\left[\left(\zeta + \beta \sqrt{\frac{(\partial \phi)^{2}}{2}} \right) R -\frac{\eta}{2} (\partial \phi)^{2} - \frac{\beta}{\sqrt{2 (\partial \phi)^{2}}}\left[(\triangle \phi)^{2}- (\nabla_{\mu} \nabla _{\nu} \phi)^{2} \right] \right],}
\end{equation}
where $ \eta$ and $\beta$ are dimensionless parameters,  they can be absorbed
 into the scalar field by means of a redefinition. The coefficient $ \zeta $ gives the Einstein- Hilbert part of the  action, which is $ \zeta = \frac{M^{2}_{pl}}{16 \pi}$.
 The  field equations  from the equation (\ref{6}) admit a static, spherically
 symmetric  and asymptotically flat solution, which is given by \cite{23},
\begin{equation}\label{7}
ds^{2} = - f(r) dt^{2} +  \frac{dr^{2}}{f(r)} + r^{2} \left( d\theta^{2} +
\sin^{2} \theta d\phi^{2}\right)
\end{equation}
where,
\begin{equation}\label{8}
f(r) = 1 -  \frac{2M}{r} - \frac{ \beta^{2} }{ 2 \zeta \eta r^{2}},
\end{equation}
where $ M $ is related to  the
black hole mass.  The parameters $\beta$ and $\eta$ should share the same sign, i,e, $ \beta > 0 $ and $\eta > 0$.\\
Here, the equation (\ref{8}) lead us to  obtain the following  temperature,
\begin{equation}\label{9}
T =\frac{ \kappa_{+}}{2 \pi}  = \frac{1}{4 \pi}\frac{df(r)}{dr}|_{r = r_{+}} =  \frac{1}{4 \pi r_{+}} \left( 1 - \frac{\gamma}{r^{2}_{+}}\right)
\end{equation}
where $ \gamma =  \frac{\beta ^{2}}{2 \zeta \eta }$.\\
Due to condition of (\ref{3}) and (\ref{5}), one can  see that the equation (\ref{9}) can not satisfied by van der Waals behavior. So, finally one can say that there was no $ P-V $ critical behavior. So, the correspondence of fluid dynamic and black hole equation of state gives us opportunity to connect between two theories from  their van der Waals behavior. When we stay in fluid dynamic side, we have van der Waals behavior. But, from black hole side there is not such equation of state. So, in that case we modify the black hole solution (\ref{8}) such that the dynamic properties of black hole coincide exactly with corresponding fluid and also such modified solution not need to additional matter to the action. So, in order to observe $ P - V$  critical behavior,  we change the Horndeski black hole with the following anstaz \cite{26},
\begin{equation}\label{10}
f(r)  = 1 -  \frac{2 M }{r} - \frac{ \gamma }{ r^{2}} + h (r, P),
\end{equation}
where the  $h(r, P)$ is function. Here,  we have to arrange such function to guarantee black hole solution with  suitable thermal properties as well  as van der Waals behavior. Such modification not change the action but it  is  satisfied by Einstein
field equations.  Also, the modification of the  above metric is a solution of the Einstein field
equations with a given energy momentum source, $G_{ab} + \Lambda g_{ab} = 8T_{ab}$.  We note here the corresponding energy-momentum source from  modified metric  should be satisfied by week, strong and dominant conditions \cite{25}. These conditions are known as energy conditions which are satisfied by our ansatz about modification of metric background.

\section{The modified Horndeski black hole}
\begin{figure}
\begin{center}$
\begin{array}{cccc}
\includegraphics[width=65 mm]{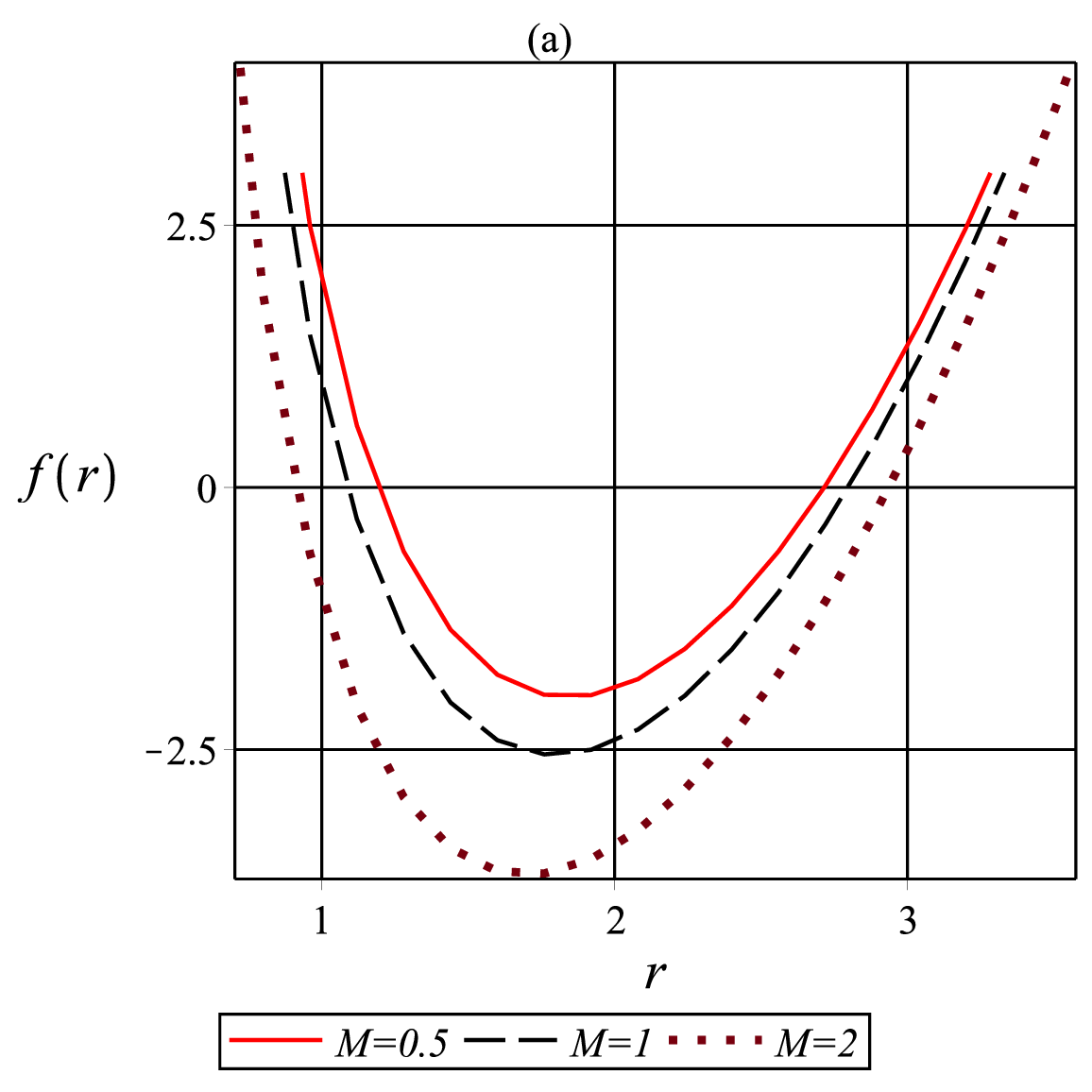}\includegraphics[width=62 mm]{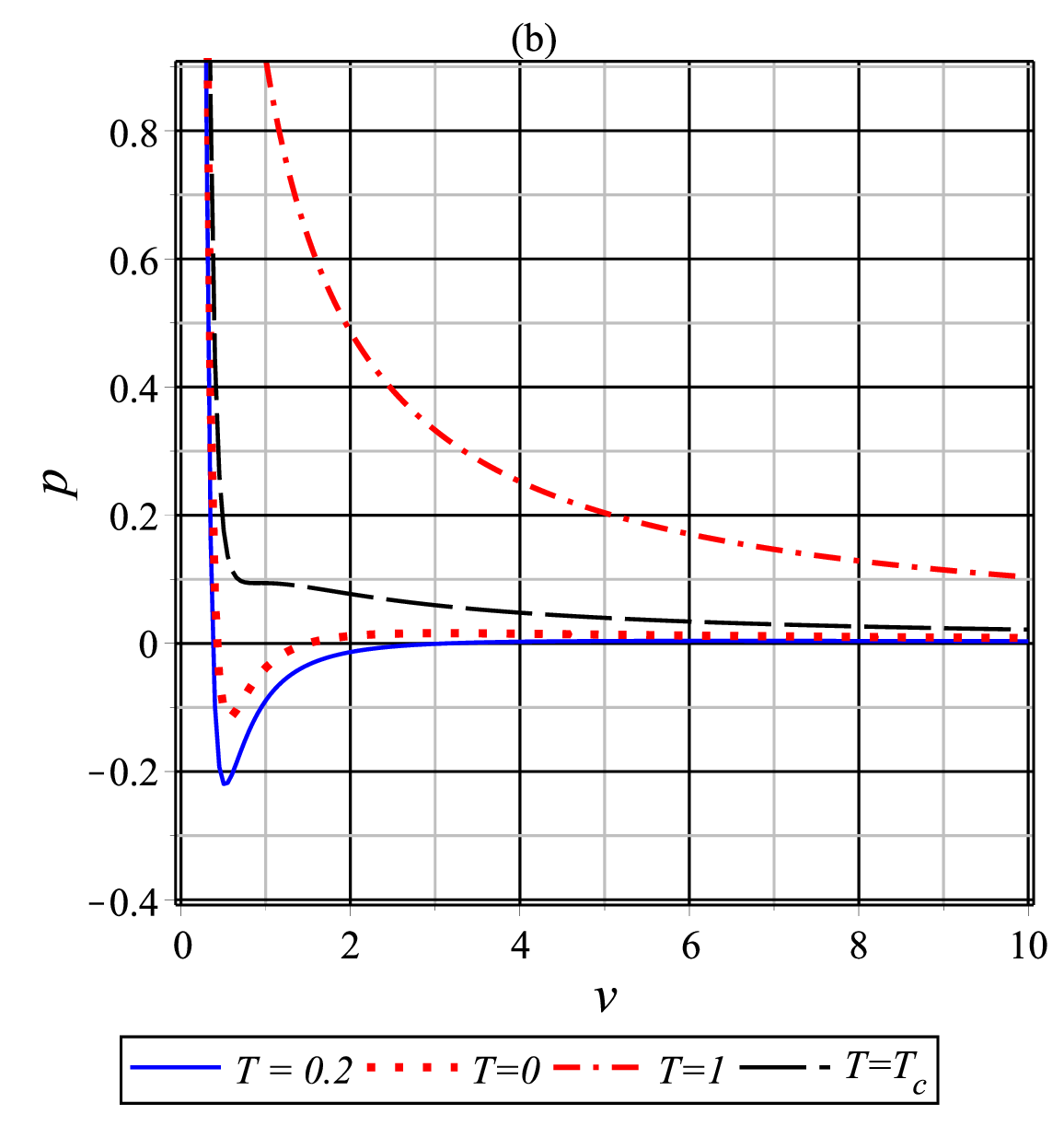}
\end{array}$
\end{center}
\caption{(a)  Horizon radius with a variation of black hole mass $ M $ and $a=1$, $b=4$, $P = 0.2$.  (b)  Pressure in terms of $\nu$ for $a=0.5$, $b=1.5$, $a=l.l$ and possible values of $T$.}
\label{fig1}
\end{figure}

In this section, we investigate the Horndeski black hole and obtain the modified metric of this black hole.
By using the Euclidean trick and equation (\ref{10}) in (\ref{9}),  one can identify the black hole temperature as \cite{26,27},
\begin{eqnarray}\label{11}
T = \frac{1}{4 \pi} \left( \frac{1}{r_{+}} + \frac{\gamma}{r^{3}_{+}} + \frac{h(r_{+},P)}{r_{+}}+  h^{\prime}(r_{+},P) \right).
\end{eqnarray}
By using the following van der Waals equation of state,
\begin{equation}\label{12}
T =  \left( P - \frac{a}{\upsilon^{2}}\right) (\upsilon + b) = P
\upsilon - P b + \frac{a}{\upsilon} - \frac{a b}{\upsilon^{2}},
\end{equation}
one can obtain $T $ as,
\begin{equation}\label{13}
T =   2 P r_{+} - P b + \frac{a}{2 r_{+}} - \frac{a b}{4 r^{2}_{+}}.
\end{equation}
where,
\begin{equation}\label{14}
\upsilon =  2 r_{+},
\end{equation}
Now, comparing (\ref{11}) and (\ref{13}) together, one can rewrite the following
expression,
\begin{equation}\label{15}
\frac{1}{4 \pi} \left( \frac{1}{r_{+}} + \frac{\gamma}{r^{3}_{+}} +
\frac{h(r_{+},P)}{r_{+}}+  h^{\prime}(r_{+},P) \right) -  2 P  r_{+} + P b
- \frac{a}{ 2 r_{+}} + \frac{a b}{4 r_{+}^{2}} = 0.
\end{equation}
In order to obtain the $ h (r_{+}, P)$,  we can rearrange  $ h (r_{+}, P)$ as
following \cite{mann, me},
\begin{eqnarray}\label{16}
h (r_{+}, P) &=& A(r_{+}) - P B(r_{+})\nonumber\\
h^{\prime} (r_{+}, P) &=& A^{\prime}(r_{+}) - P B^{\prime}(r_{+}).
\end{eqnarray}
From (\ref{15}) and (\ref{16}),  one can obtain the following expression,
\begin{equation}\label{17}
P \left( b - 2 r_{+} - \frac{B(r_{+}) }{ 4 \pi r_{+}} -
\frac{B^{\prime}(r_{+}) }{ 4 \pi } \right) - \left(\frac{a}{2 r_{+}} -
\frac{a b}{4 r^{2}_{+}} - \frac{1}{4 \pi r_{+}} - \frac{\gamma}{4
\pi r^{3}_{+}} - \frac{A(r_{+})}{4 \pi r_{+}} - \frac{A^{\prime}(r_{+})}{4
\pi } \right) = 0.
\end{equation}
Here, two terms must be independently zero. So ,we have,
\begin{equation}\label{18}
b - 2 r_{+} - \frac{B(r_{+}) }{ 4 \pi r_{+}} -
\frac{B^{\prime}(r_{+}) }{ 4 \pi } = 0,
\end{equation}
one can obtain $ B(r_{+})$ as,
\begin{equation}\label{19}
B(r_{+}) = 4 \pi \left( b \frac{r_{+}}{2} - \frac{2}{3} r^{2}_{+}
\right).
\end{equation}                                                                                                                                                                                                                                                                                                                                                                                                                                                     Again, the  second term will be,
\begin{equation}\label{20}
A^{\prime}(r_{+}) + \frac{A(r_{+})}{r_{+}} = \frac{( 2 \pi a -1 )}{r_{+}} -
\frac{\pi a b}{r^{2}_{+}} - \frac{\gamma}{r^{3}_{+}},
\end{equation}
 and  $ A (r_{+})$ is given by the following equation,
\begin{equation}\label{21}
A(r_{+}) = ( 2 \pi a -1 ) - \pi a b \frac{\ln(r_{+})}{r_{+}} +
\frac{\gamma}{r^{2}_{+}}.
\end{equation}
So, from the previous anstaz,  one can obtain $ h( r_{+}, p)$ as follows,
\begin{equation}\label{22}
h( r_{+}, p ) = ( 2 \pi a -1 ) - \pi a b \frac{\ln(r_{+})}{r_{+}} +
\frac{\gamma}{r^{2}_{+}} + \frac{2}{3} \pi P \left( 4 r^{2}_{+} - 3 b r_{+}  \right).
\end{equation}
So, the modified metric function is given by,
\begin{equation}\label{23}
f(r_{+}) = 2 \pi a  - \frac{2M}{r_{+}} - \pi a b
\frac{\ln(r_{+})}{r_{+}}  + \frac{2}{3} \pi P \left( 4 r^{2}_{+} - 3 b r_{+}  \right).
\end{equation}
So, here we  modified Horndeski black hole by new definition of $h(r_{+},P)$ anstaz.
We plot in Fig. \ref{fig1} (a)  the function $ f ( r_{+}) $ in terms of horizon radius for the corresponding  black hole.  Here, one can see that there exists a critical point for $ f (r_{+}) $ which decreases as long as $ M $ mass of the black hole increases in the critical point. It is clear from black dashed line of the Fig. \ref{fig1} (b). By increasing the physical mass, the function  $ f ( r_{+}) $  decreases and increases before and after the critical horizon radius respectively.

In the next sections, we use of equation (\ref{23}) and  obtain the $ P - V $ critically  points, in that case we employ the ordinary and the new method.

\section{Ordinary approach to the modified Horndeski  black hole}
\begin{figure}
\begin{center}$
\begin{array}{cccc}
\includegraphics[width=55 mm]{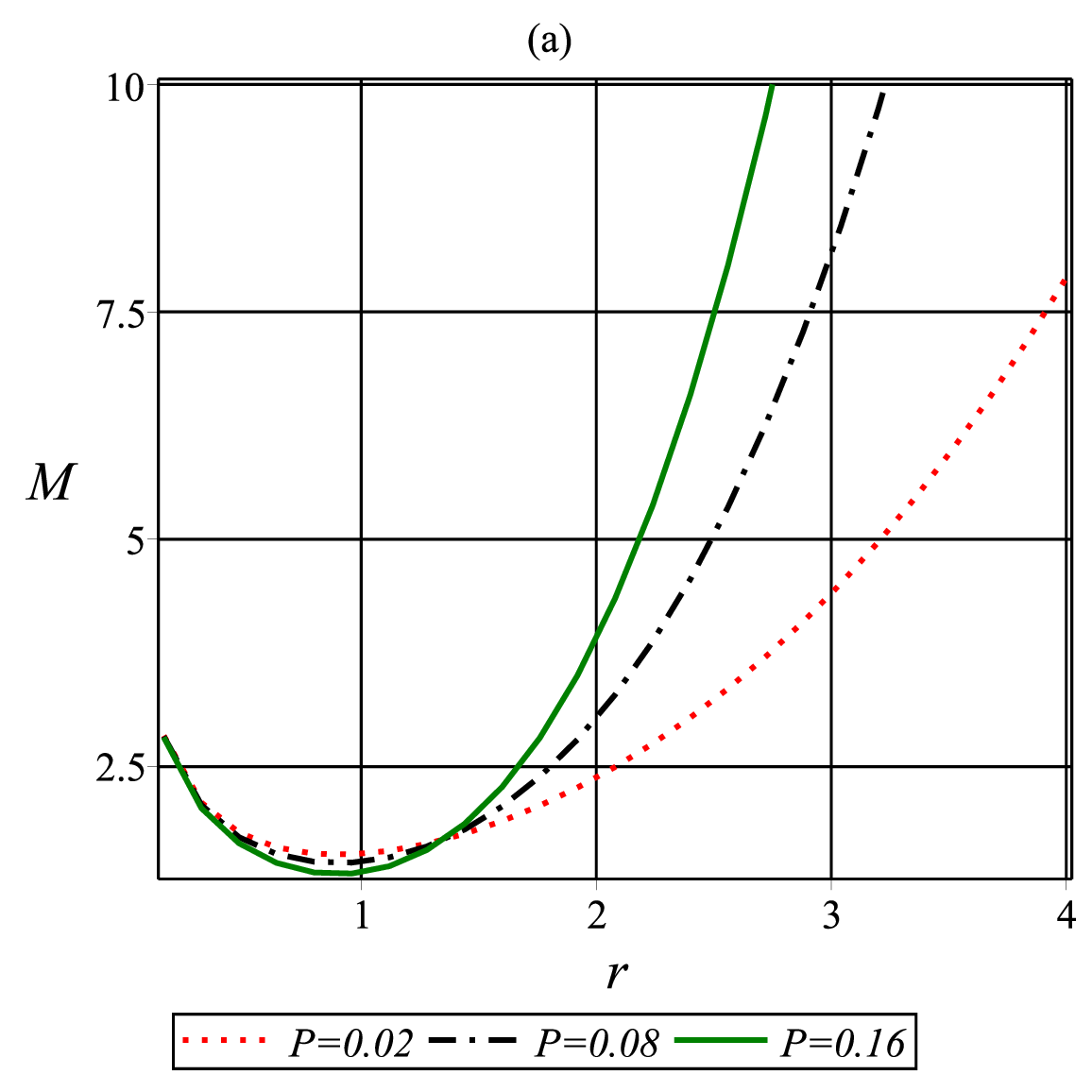}\includegraphics[width=55 mm]{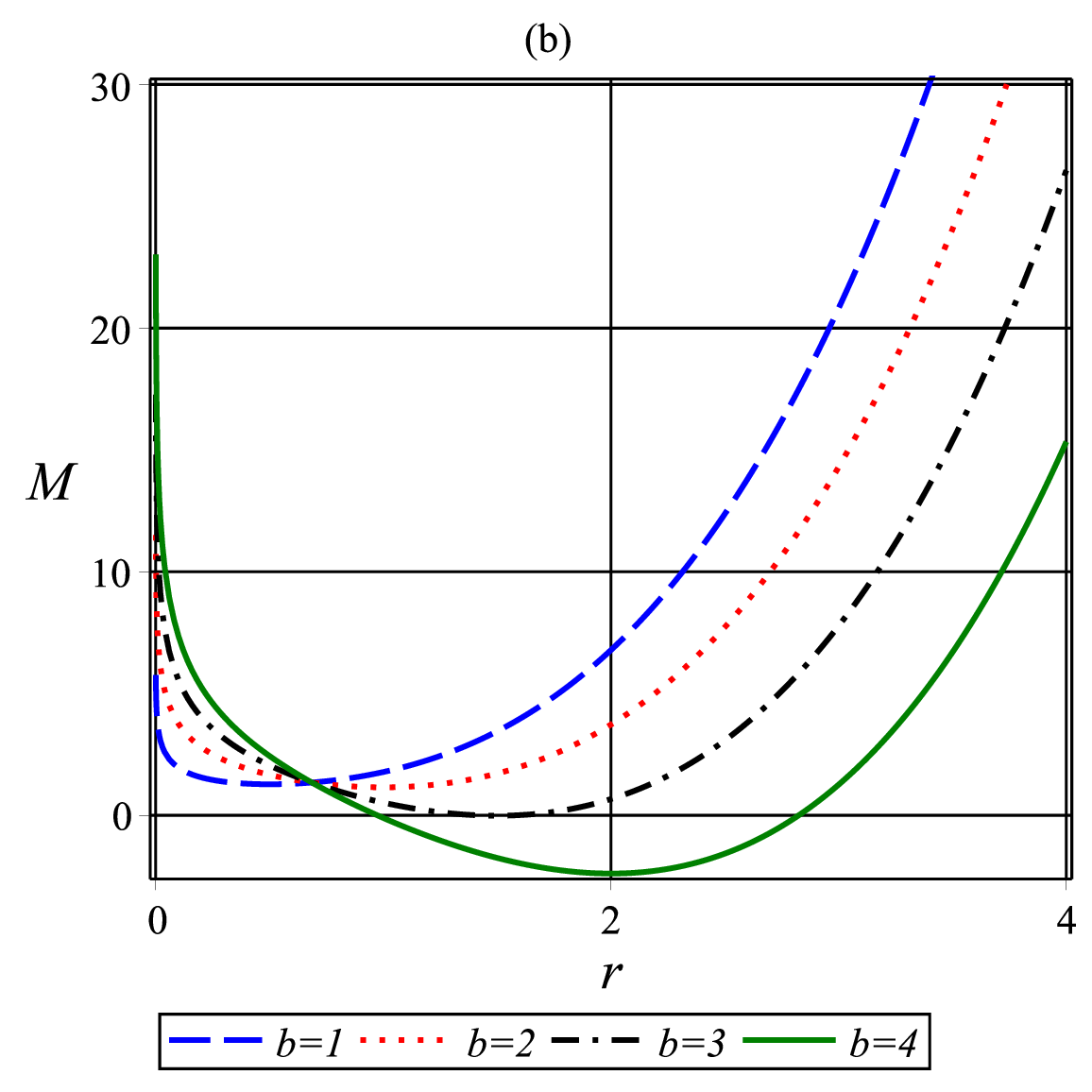}\includegraphics[width=55 mm]{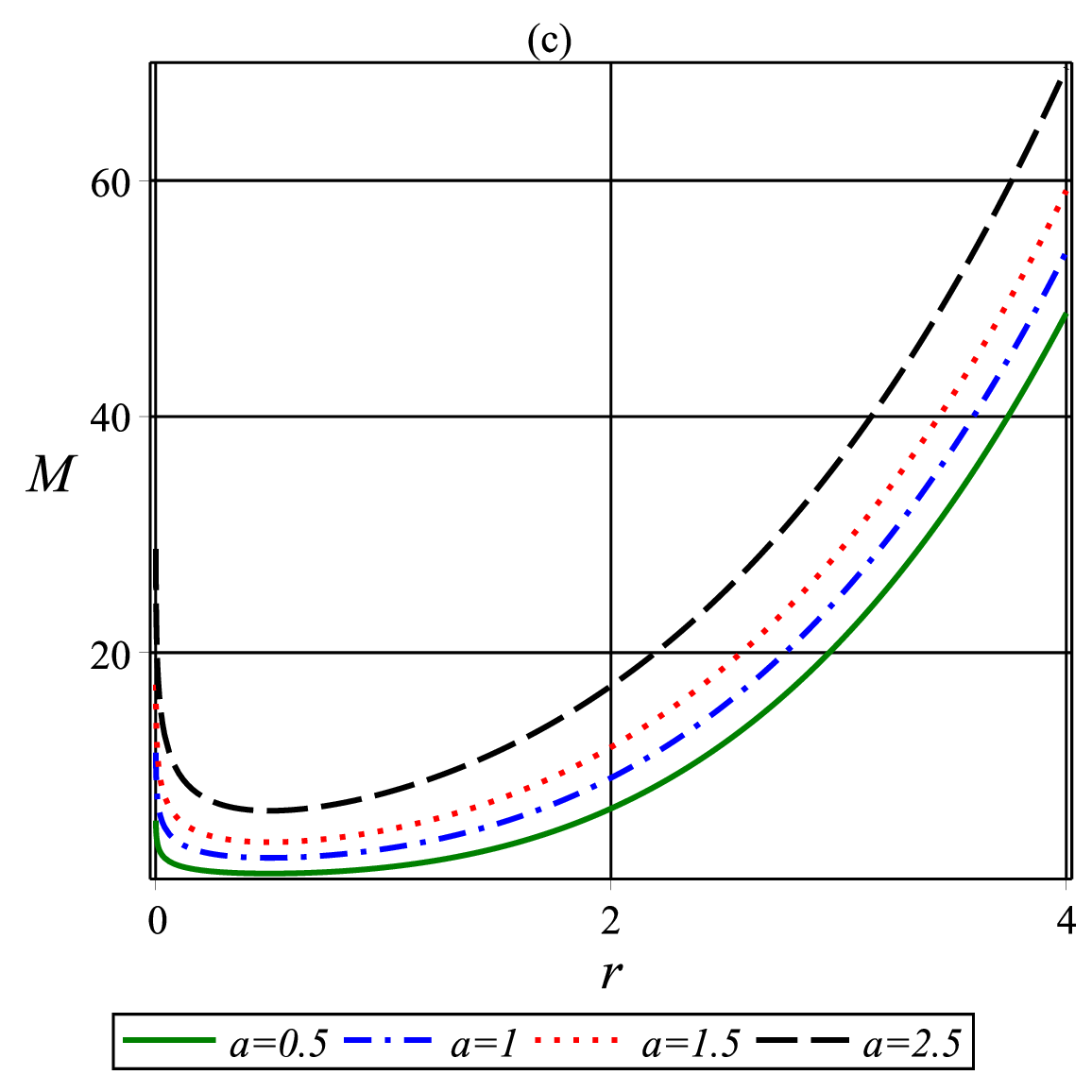}
\end{array}$
\end{center}
\caption{The physical mass in terms of horizon radius  for, (a) $a=0.5$, $ b=1.8$ and all possible values of $P$, (b)  $a=0.5$,  $ P=0.2$ and all possible values of $b$, (c) $b=1$, $ P=0.2$ and all possible values of $a$.}
\label{fig2}
\end{figure}

Now, we are going to investigate the $ P - V $ critically for the modified metric. In that case, we
need the some thermodynamical quantities which play an important role for the study of the $ P - V $ critically system.
As we know, the position of the black hole event horizon is determined with $ f(r_{+}) = 0. $ The parameter $ M $
represent the ADM mass of the black hole, where the physical mass is given by,
\begin{equation}\label{24}
M = \pi a r_{+} -   \frac{\pi a b }{2 } \ln {r_{+}} + 2 \pi P (\frac{2 r^{3}_{+}}{3} - b\frac{r^{2}_{+}}{2} ).
\end{equation}
We see in  Fig. \ref{fig2}   the behavior of  physical mass in terms of horizon radius.\\

Here, one can see that  the values of physical mass  decreases and increases before and after the minimum point respectively.\\
Also, we see that there exists a critical point for physical mass which decreases as well as increasing
the coefficients $ a $ and $ b $. By increasing parameter $ P $ the physical mass increases and decreases
before and after the critical point respectively.\\
Also, we have to  calculate  the Hawking temperature  as following,
\begin{equation}\label{25}
T =  \frac{a}{2 r_{+} } - \frac{a b }{4 r^{2}_{+}} + P (2 r_{+} - b ).
\end{equation}
The  black hole entropy is given by,
\begin{equation}\label{26}
S = \frac{\mathcal{A}}{4} = 4 \pi  r^{2}_{+}.
\end{equation}

\begin{figure}
\begin{center}$
\begin{array}{cccc}
\includegraphics[width=55 mm]{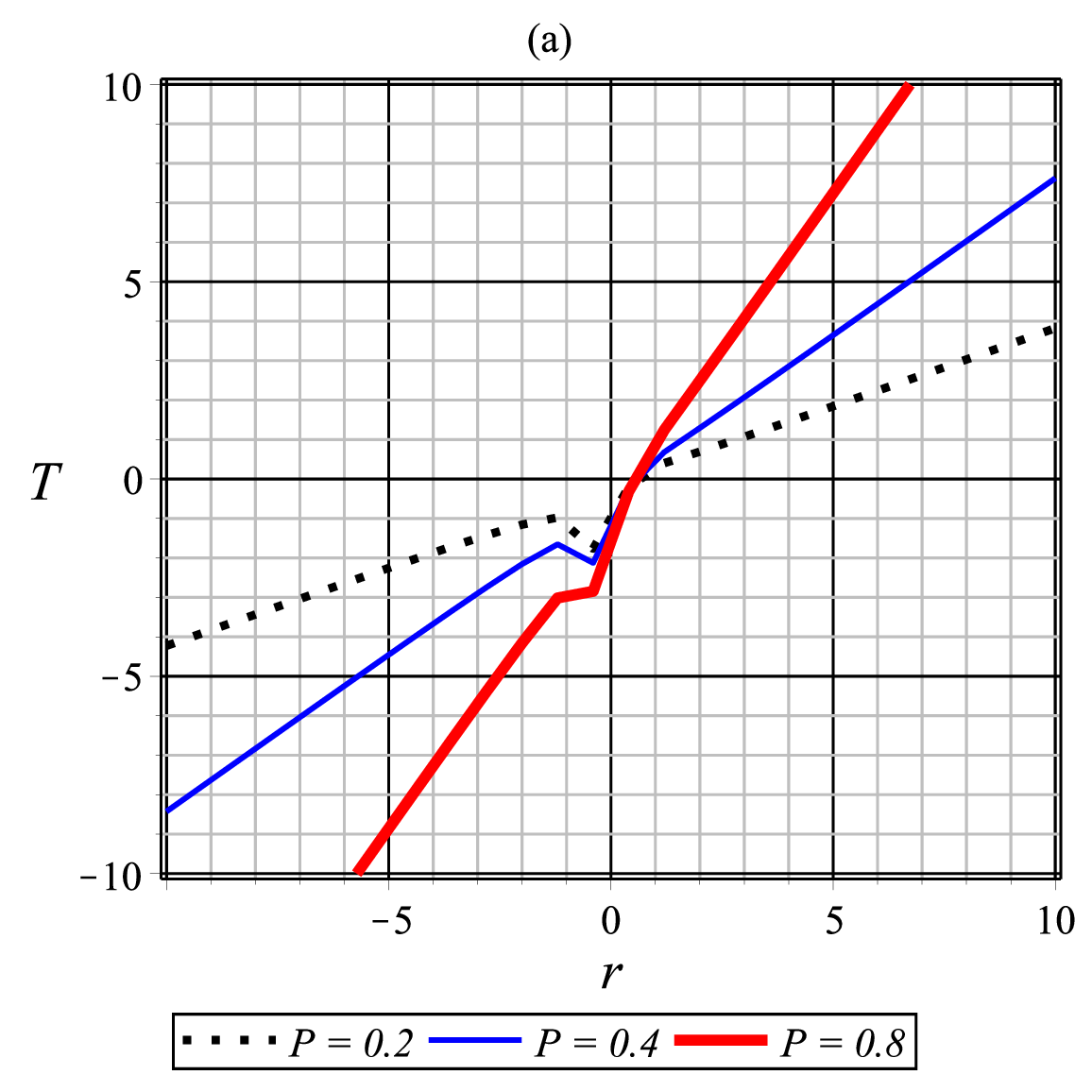}\includegraphics[width=55 mm]{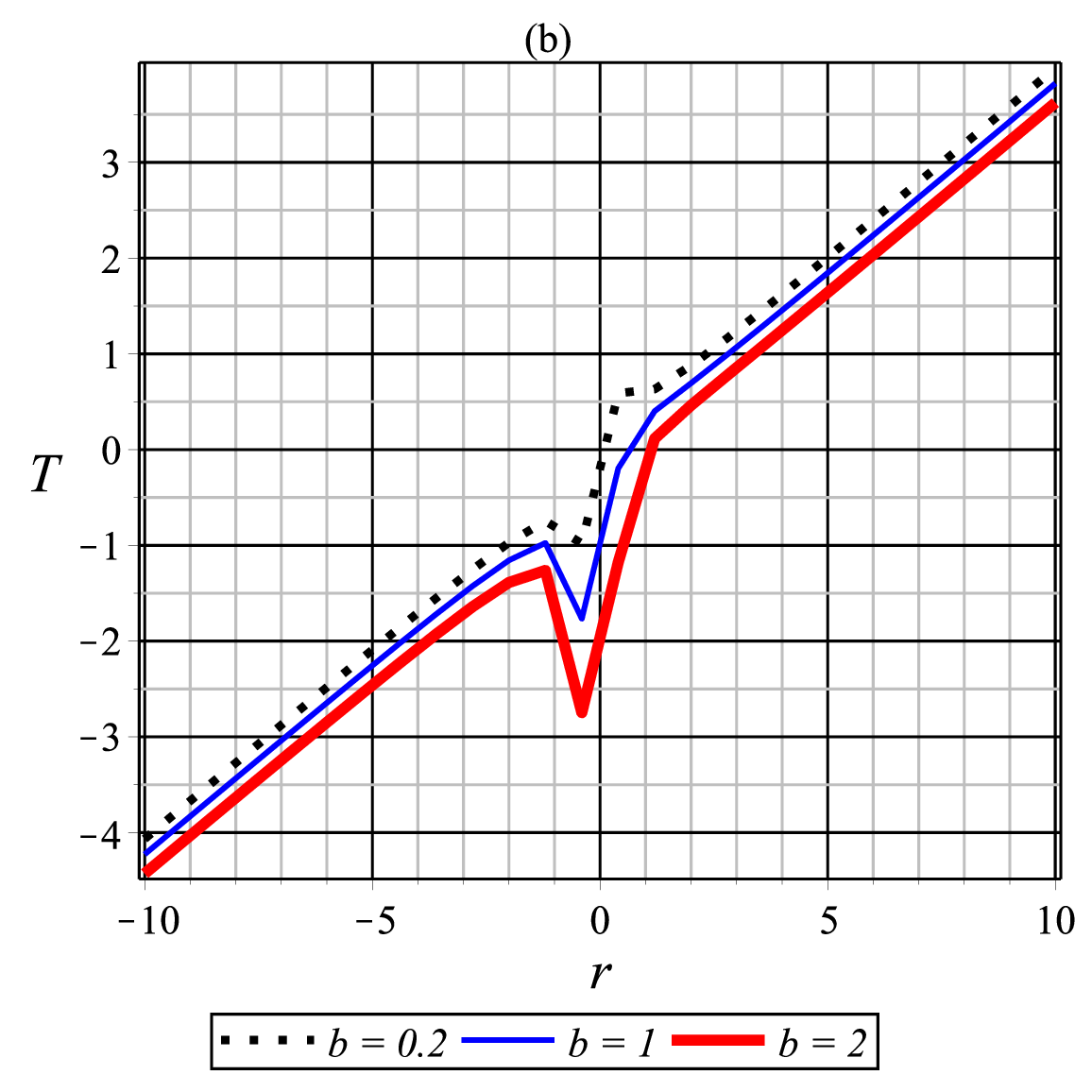}\includegraphics[width=55 mm]{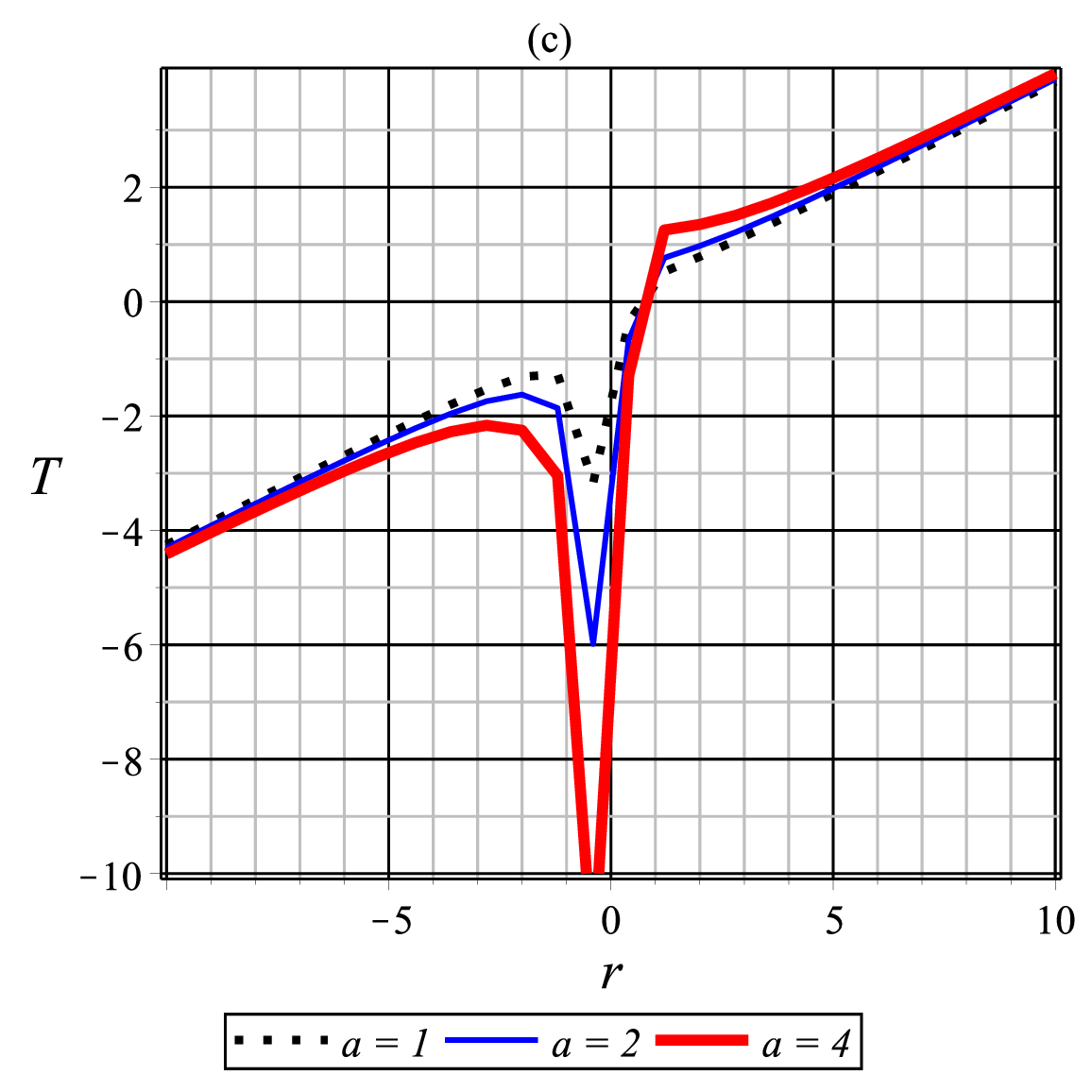}
\end{array}$
\end{center}
\caption{The temperature in terms of horizon radius  for, (a)  $a=0.5$, $ b=1$ and all possible values of $P$, (b)  $a=0.5$, $ P=0.2$ and all possible values of $b$, (c)  $b=1$, $ P=0.2$ and all possible values of $a$.}
\label{fig3}
\end{figure}

We plot in  Fig. \ref{fig3}  the behavior of the temperature in terms of  horizon radius.
Here, one can see that  there exists a critical point for temperature which  decreases and increases before and after the critical horizon radius respectively.                                                                                                                                                                                                                                                   Also, the temperature has decreases when coefficients of van der Waals are  increases.
As we know, the first - order phase transition occurs when the temperature is zero. In that case, we observe the  first - order  phase transition in the points of $  r_{+} = 0.42$ and $  r_{+} = 1.71$.
The phase transition of type two is associated with divergence points of the specific heat. It means that the singular points of the specific heat are representing the phase transition.\\
The pressure quantity is given by,
\begin{equation}\label{27}
P = \frac{1}{\upsilon - b} \left( T - \frac{a}{\upsilon} + \frac{a b}{\upsilon^{2}}\right).
\end{equation}

We can use the equation (\ref{27}) to investigate the $ P - V $ diagram of a the modified Horndeski black hole.
Now, we have  following conditions and obtain the critical points for the $ T_{c}$, $P_{c},$ and $\nu_{c}$.\\
\begin{eqnarray}\label{28}
\frac{\partial P}{\partial \upsilon } = 0,\nonumber\\
\frac{\partial^{2} P}{\partial \upsilon^{2} } = 0.
\end{eqnarray}
Therefore, by using the equations (\ref{27}) and (\ref{28}), one can obtain the following critical points for
$P_{c}$, $\upsilon_{c}$, and $T_{c}$,
\begin{eqnarray}\label{29}
P_{c} &=& \frac{a}{27 b^{2}},\nonumber\\
T_{c} &=&  \frac{8 a}{ 27 b},\nonumber\\
\upsilon_{c} &=& 3 b.
\end{eqnarray}
So, we use equation (\ref{29}), one can obtain following expression,
\begin{equation}\label{30}
\rho = \frac{P_{c} \upsilon_{c}}{ T_{c}} = \frac{3}{8},
\end{equation}
where $\rho$ is a universal constant in ideal gas.
The above product is equal to $ \frac{3 }{8} $,
in that case we find an interesting relation which is exactly the
same as the van der Walls fluid, and it is a universal number
predicted for the modified Horndeski black hole.\\
The typical behavior of the $ P - V $ diagram corresponding to the modified Horndeski black hole is plotted in Fig. \ref{fig1} (b).
 We can see black dashed lines in Fig. \ref{fig1} (b) to find that a modified Horndeski black hole is also the dual of van der Waals fluid.

\subsection{Stability and Phase Transition }
As we know there are several methods to investigate the thermal stability and phase transition. We need  quantities which play important role for the study of stability system as Gibbs free energy and heat capacity. When the Gibbs free energy is negative $ (G < 0)$, the system has a global stability. Also, when it is positive $ (G > 0)$, the system has a local stability.  In order to discuss the global and local stability of the black hole we need to calculate the Gibbs free energy  which is given by,
\begin{equation}\label{31}
G = M - ST =  \pi a b - \pi a r_{+} -   \frac{\pi a b }{2 } \ln {r_{+}} -  \pi P r^{2}_{+} \left(\frac{8}{3} r_{+} - b \right),
\end{equation}

\begin{figure}
\begin{center}$
\begin{array}{cccc}
\includegraphics[width=55 mm]{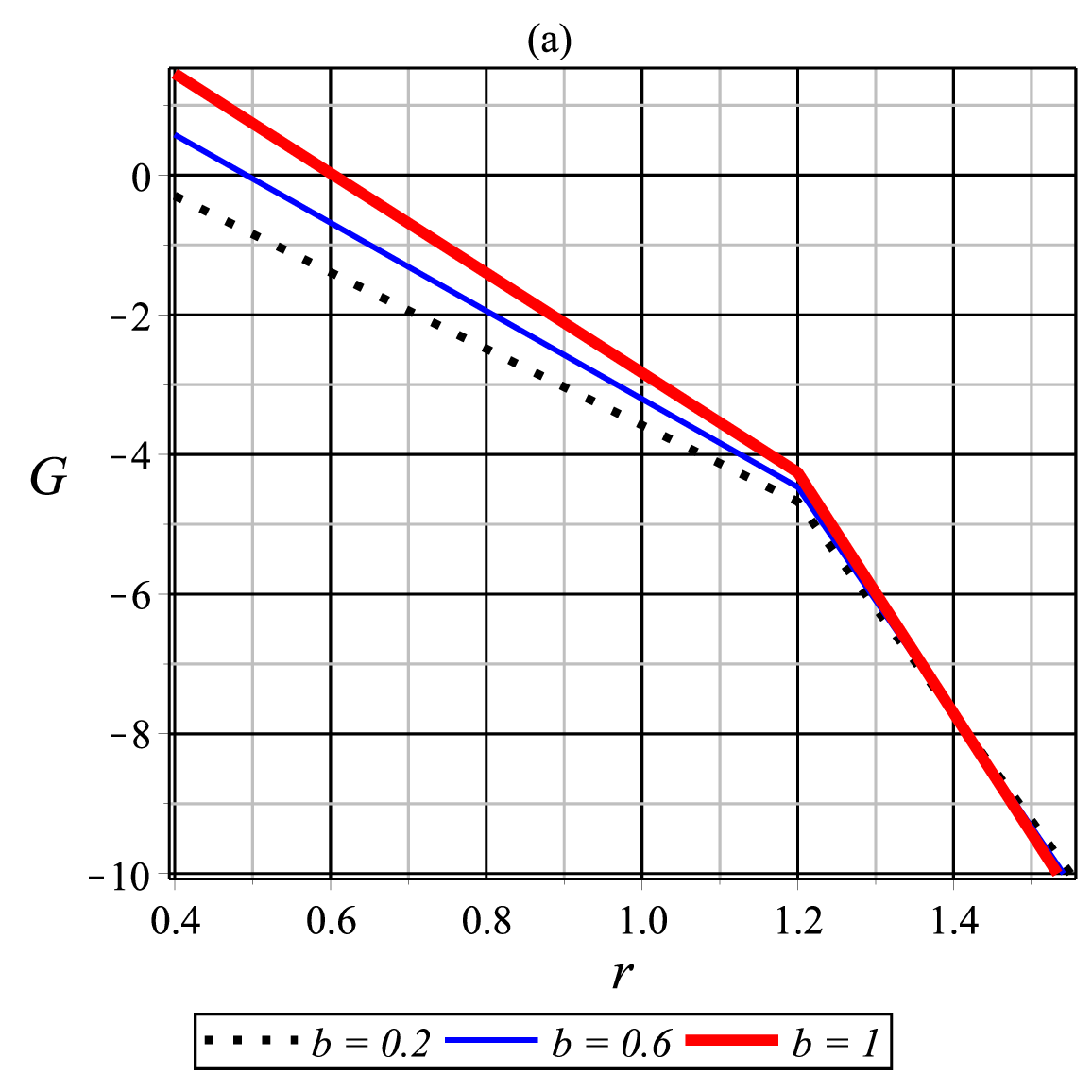}\includegraphics[width=55 mm]{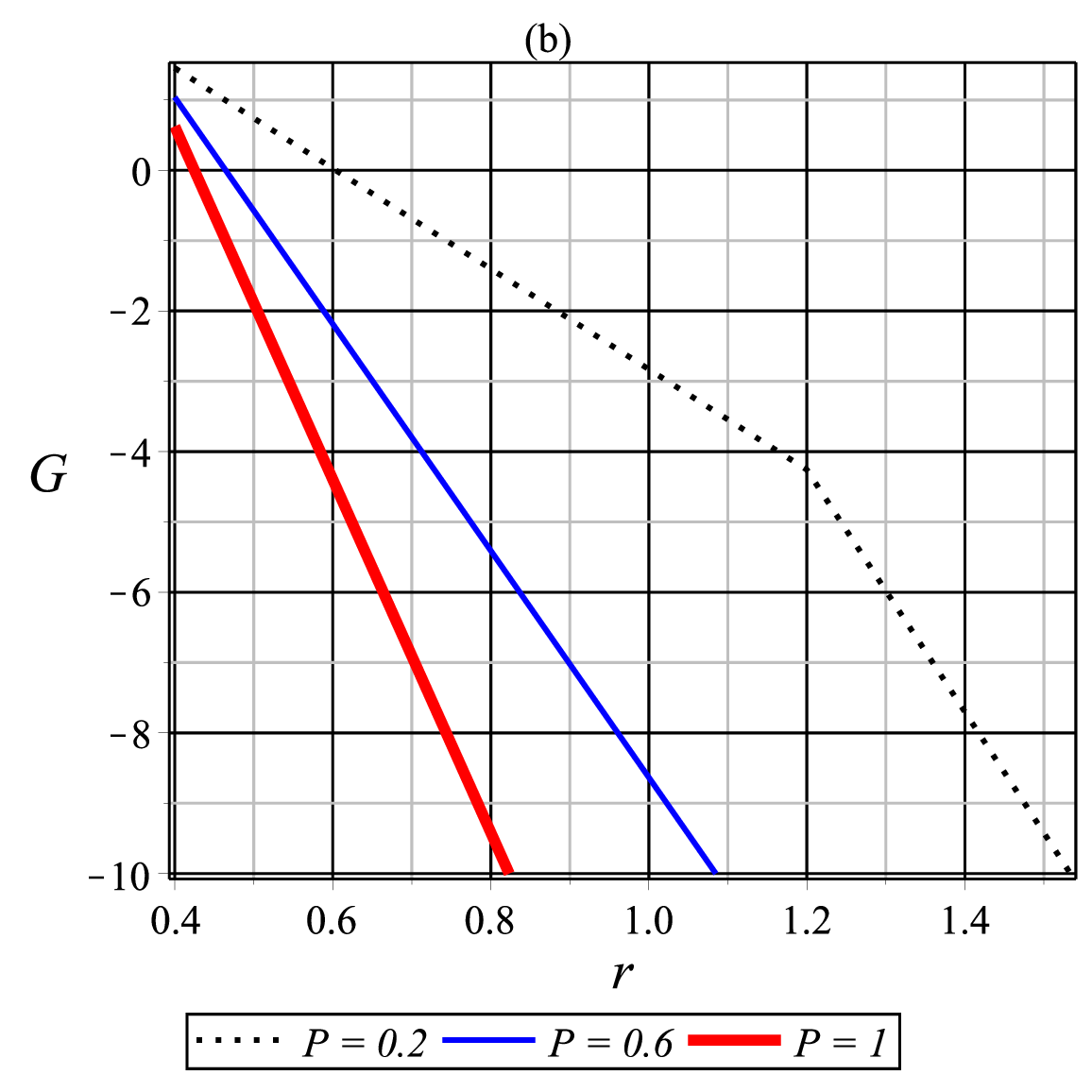}\includegraphics[width=55 mm]{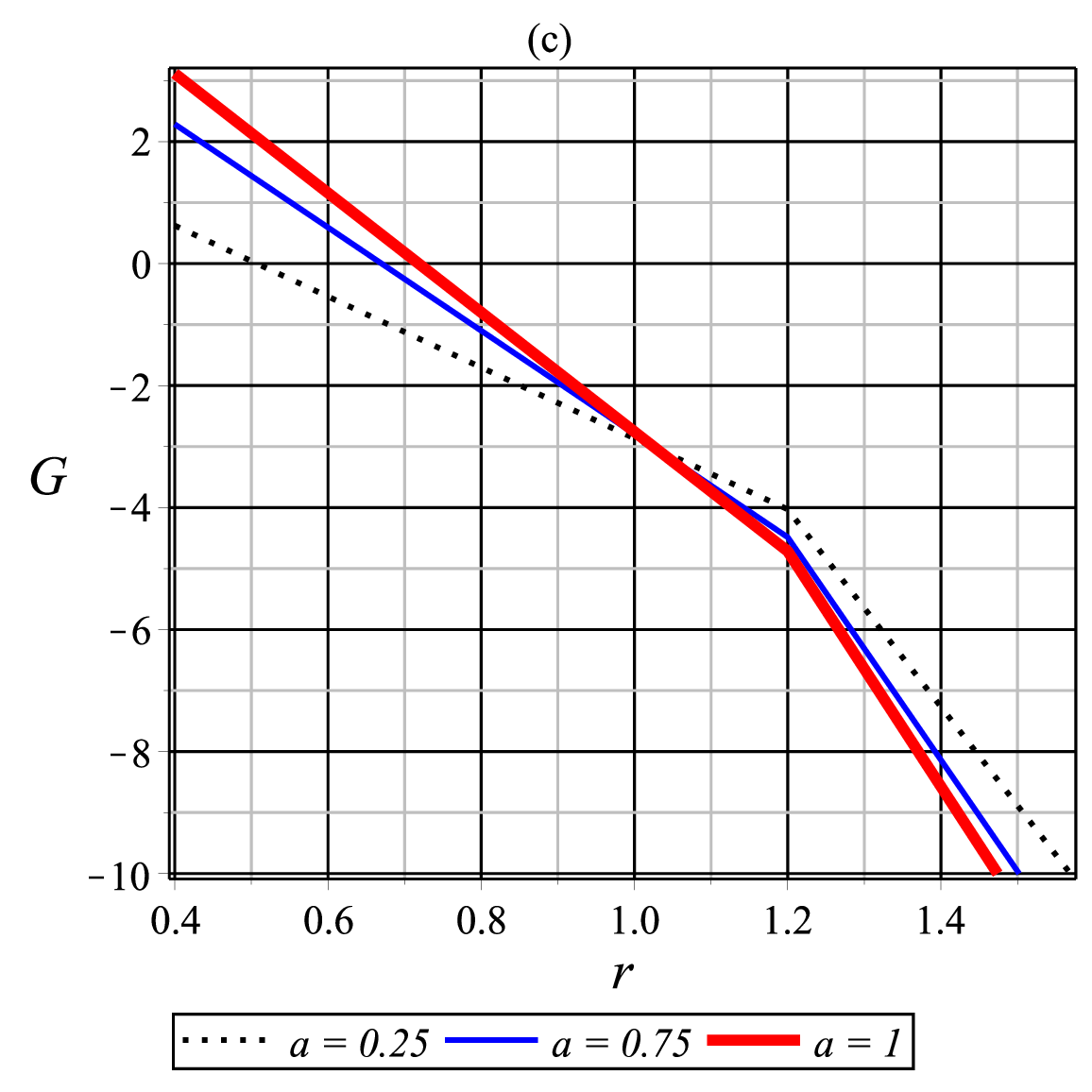}
\end{array}$
\end{center}
\caption{The Gibbs free energy for; (a) all possible  values of $b$, (b) all possible values of $P$ , (c) all possible  values of $a$.}
\label{fig4}
\end{figure}

The graphical analysis of the  Gibbs free energy for modified Horndeski  black hole can be seen in Fig. \ref{fig4}.  We observe in Figs. \ref{fig4}, by increasing horizon radius, the Gibbs free energy has global stability.  Also, it has local stability in the very small  ranges of  horizon radius.
By increasing the parameter of $P$ the Gibbs free energy decreases. So,  the system is quite global stability.
For different  values of  parameters $a$ and $b$  there exists a critical point for  the Gibbs free energy which  increases and decreases before and after the critical point  respectively. Also, we observe that for the positive horizon radius the Gibbs free energy is unstable.\\

As we know, the specific heat is an important measurable physical quantity which can determine the stability of the system.
Here, we study two types of phase transition as type one and type two.\\
As we discussed in the previous section, the phase transition of type one occurs when the temperature vanishes.  So, to putting  $T = 0$ in equation (\ref{27}) we will have,
\begin{equation}\label{32}
P|_{T = 0 } = - \frac{a}{4  r^{2}_{+}}.
\end{equation}
It means that $ T = 0$ indicate a bound point between nonphysical $ (T < 0)$ and physical $ (T > 0)$ regimes.\\
Also, the specific heat has the following relation with mass $ M $, entropy $S$, and temperature $T$,
\begin{equation}\label{33}
C = T \left (\frac{\partial S}{\partial T} \right ) = \frac {\left (\frac{\partial M}{\partial S} \right )}{\left (\frac{\partial^{2} M}{\partial S^{2}} \right )} .
\end{equation}
By using the equations (\ref{25}) and (\ref{26}), we  obtain the specific heat as,
\begin{equation}\label{34}
C = \left (\frac{2 a \sqrt{\pi}S^{\frac{3}{2}}- \pi a b S + 4 S^{2} P (\frac{2}{\sqrt{\pi}}S^{\frac{1}{2}} - b)}{- a \sqrt{\pi}S^{\frac{1}{2}}+ \pi a b + \frac{P}{\sqrt{\pi}}S^{\frac{3}{2}}} \right ).
\end{equation}
If $ C > 0 $, the black hole is in stable state, and if $C < 0$, the black hole is in unstable state. As regards the change of sign in specific heat with asymptotic behavior is representing the phase transition between unstable / stable states.  So, $ C = 0$ corresponds to the phase transition of the van der Walls fluid similar to the critical point discussed above which leads to the following equations,
\begin{eqnarray}\label{35}
2 a \sqrt{\pi}S^{\frac{3}{2}}- \pi a b S + 4 S^{2} P (\frac{2}{\sqrt{\pi}}S^{\frac{1}{2}} - b)= 0,\nonumber\\
2 a r_{+} - a b + 4 P r^{2}_{+}(2 r_{+} - b) = (a + 4 r^{2}_{+} P )(2 r_{+} - b)= 0.
\end{eqnarray}
That this agree with the equation (\ref{32}). As we said, the phase transition of type two is associated with divergence points of the specific heat. It means that $\frac{\partial^{2} M}{\partial S^{2}} = 0$. Therefore,
\begin{equation}\label{36}
8 P r_{+}^{3} - 2 a r_{+}+  a b = 0
\end{equation}
which yields to the following root,
$$  r_{+} = \frac{0.18 D^{\frac{2}{3}} + 0.21 a P}{P D^{\frac{1}{3}}}.$$
where $ D =\left[  a (-1.12 b + 0.2 \sqrt{\frac{-4 a + 27 b^2 P}{P}})P^{2}\right].$\\
Later, we discuss about specific heat and phase transition, and compare with results obtained by other method.

\section{New approach to the modified Horndeski  black hole}
Regarding the review of ordinary thermodynamic systems, it is evident that all of the complete differentiations  can be written as a function of three thermodynamic coordinates.
 These three coordinates are not independent, for example, in most cases, thermodynamic systems can be written in terms of pressure, temperature, internal energy, and free energy of Gibbs, which are independent of each other.
We want to get relationships that are independent of each other, but these new relationships must satisfy conditions related to the thermodynamic behavior of the system, such as phase transition and critical behavior.
So, instead of using the equation of the ordinary state (which is temperature dependent), we obtain the slope of the temperature in terms  the entropy. This equation gives us a new relationship that involves pressure, which is only depend on volume.\\
Now, we want to present a new method to present these relations for different thermodynamic variables. These new relationships provide the conditions for the system phase transition.
We obtain the slope of the temperature in terms of the entropy. This equation gives us a new relationship that involves pressure, which is only depend on the volume. It should be noted that in order to obtain a new relationship for pressure, one can use $ \left( \frac{\partial^{2} H}{\partial s^{2} }\right) $ instead of $ \frac{\partial T}{\partial S} $, where $ H $ is an enthalpy of the system.\\

Now, we calculate the volume conjugating to the pressure,
\begin{equation}\label{37}
V = \left( \frac{\partial H}{\partial P }\right)_{s} =  \left( \frac{\partial M}{\partial P }\right)_{s} = \frac{4 \pi}{3}r^{3}_{+} - \pi b r^{2}_{+},
\end{equation}
where the black hole mass considered as the black hole enthalpy \cite{Dolan, HL}. Thus, we are in a position to use the new method. Since both  entropy $ S (\nu )$ and enthalpy $ H ( \nu)$ are volume dependent, we can use the following relation \cite{22},
\begin{equation}\label{38}
\left( \frac{\partial H}{\partial S }\right)_{Q} = a \frac{\sqrt{\pi}}{2} S^{\frac{-1}{2}} - \frac{ \pi a b }{4 S} + 2 \frac{P \sqrt{S}}{\sqrt{\pi}} - Pb
\end{equation}
and,
\begin{equation}\label{39}
\left( \frac{\partial^{2} H}{\partial S^{2} }\right)_{Q} = \frac{1}{S} \left(- \frac{a \sqrt{\pi}}{4} S^{\frac{-1}{2}} + \frac{ \pi a b }{4 S} + \frac{P \sqrt{S}}{\sqrt{\pi}} \right ).
\end{equation}
In order to solve this relation with respect to $ P $, one can find following new relation for pressure which differs from equation of state,
\begin{equation}\label{40}
P_{new} =  \frac{a}{ 16 r^{2}_{+}}  - \frac{ a b }{ 32 r^{3}_{+}}.
\end{equation}
Using the concept of extremum of this relation being the critical point, one can find critical volume and pressure,
\begin{eqnarray}\label{41}
\upsilon_{c} &=& 3 b,\nonumber\\
P_{c} &=& \frac{a}{54 b^{2}}
\end{eqnarray}
Regarding this relation and replacing corresponding pressure in the temperature (\ref{24}), mass (\ref{27}) and Gibbs free energy (\ref{31}), we obtain  the new relation  as following,
\begin{equation}\label{42}
T_{new} = \frac{a}{ r_{+} } - \frac{a b }{ r^{2}_{+}} + \frac{a b^{2}}{4 r^{3} _{+}},
\end{equation}

\begin{figure}
\begin{center}$
\begin{array}{cccc}
\includegraphics[width=60 mm]{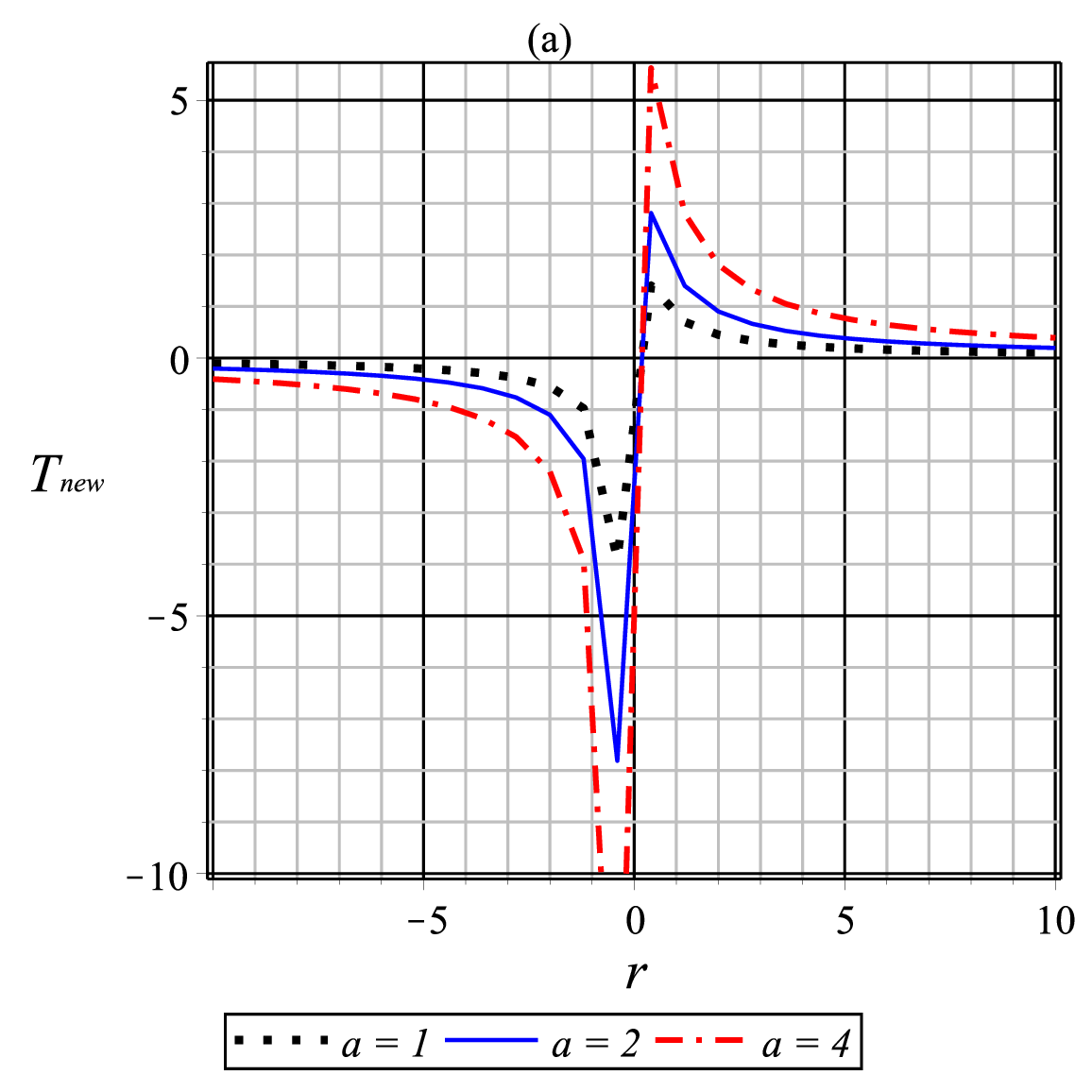}\includegraphics[width=60 mm]{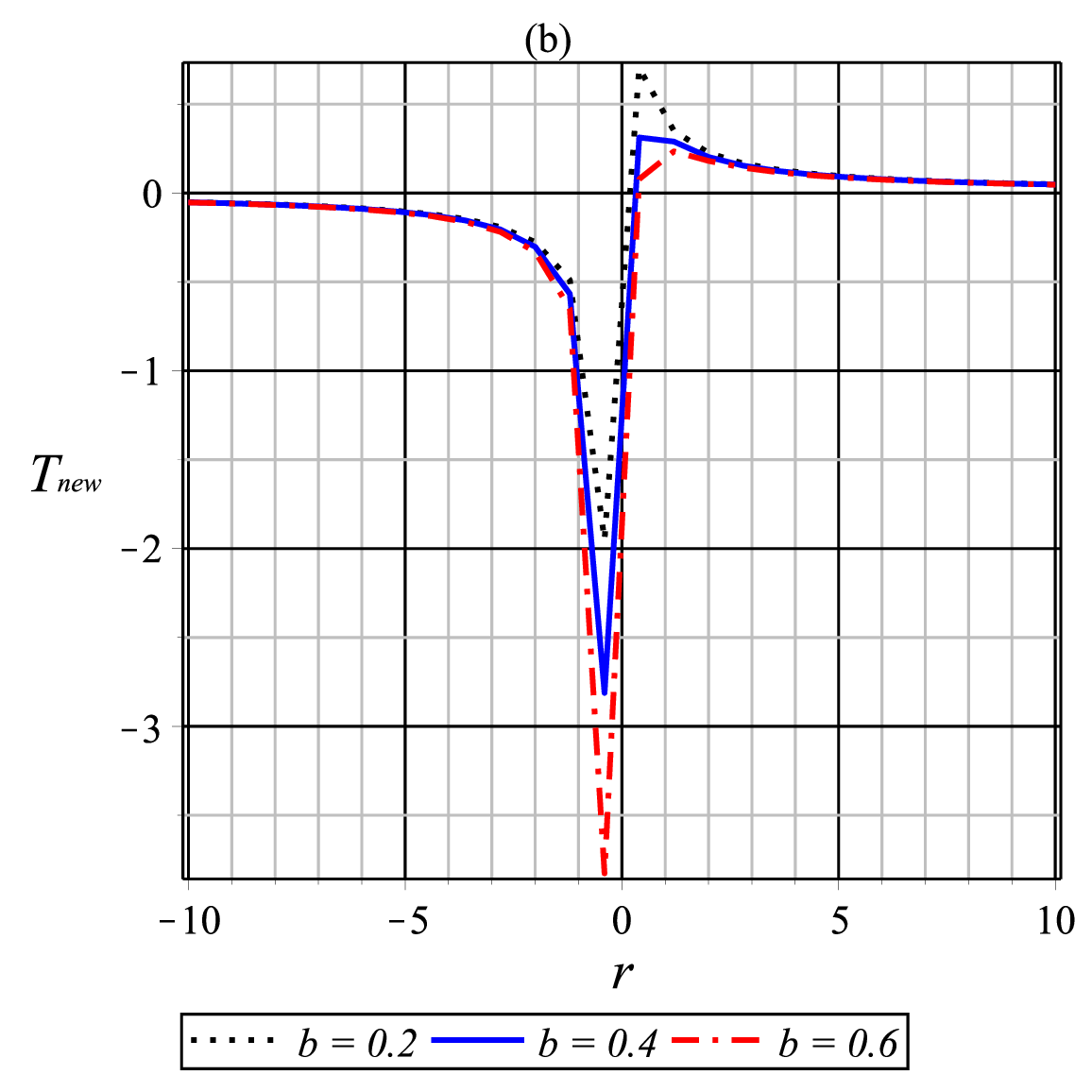}
\end{array}$
\end{center}
\caption{The new temperature;  (a) $b = 1$, and all possible values of $a$, (b)  $a=0.5$, and all possible values of $b$.}
\label{fig5}
\end{figure}

The effect of quantum variables on the temperature  in terms of horizon radius can be seen in Fig. \ref{fig5}.  We can observe  that the effects quantum variables  established for small black holes.
It can be seen that for small  values of such  parameters  the behavior temperature decreasing and increasing  in critical point respectively.
To compare Fig. \ref{fig3} and Fig. \ref{fig5},  one can see that  there are two critical points for the  new temperature diagram.

Here, the new mass is given by,
\begin{equation}\label{43}
M_{new} = \frac{4}{3} \pi a r_{+} -   \frac{\pi a b }{2 } \ln {r_{+}} - \frac{7}{12} \pi a b + \frac{\pi a b^{2}}{4 r_{+}}.
\end{equation}

\begin{figure}
\begin{center}$
\begin{array}{cccc}
\includegraphics[width=60 mm]{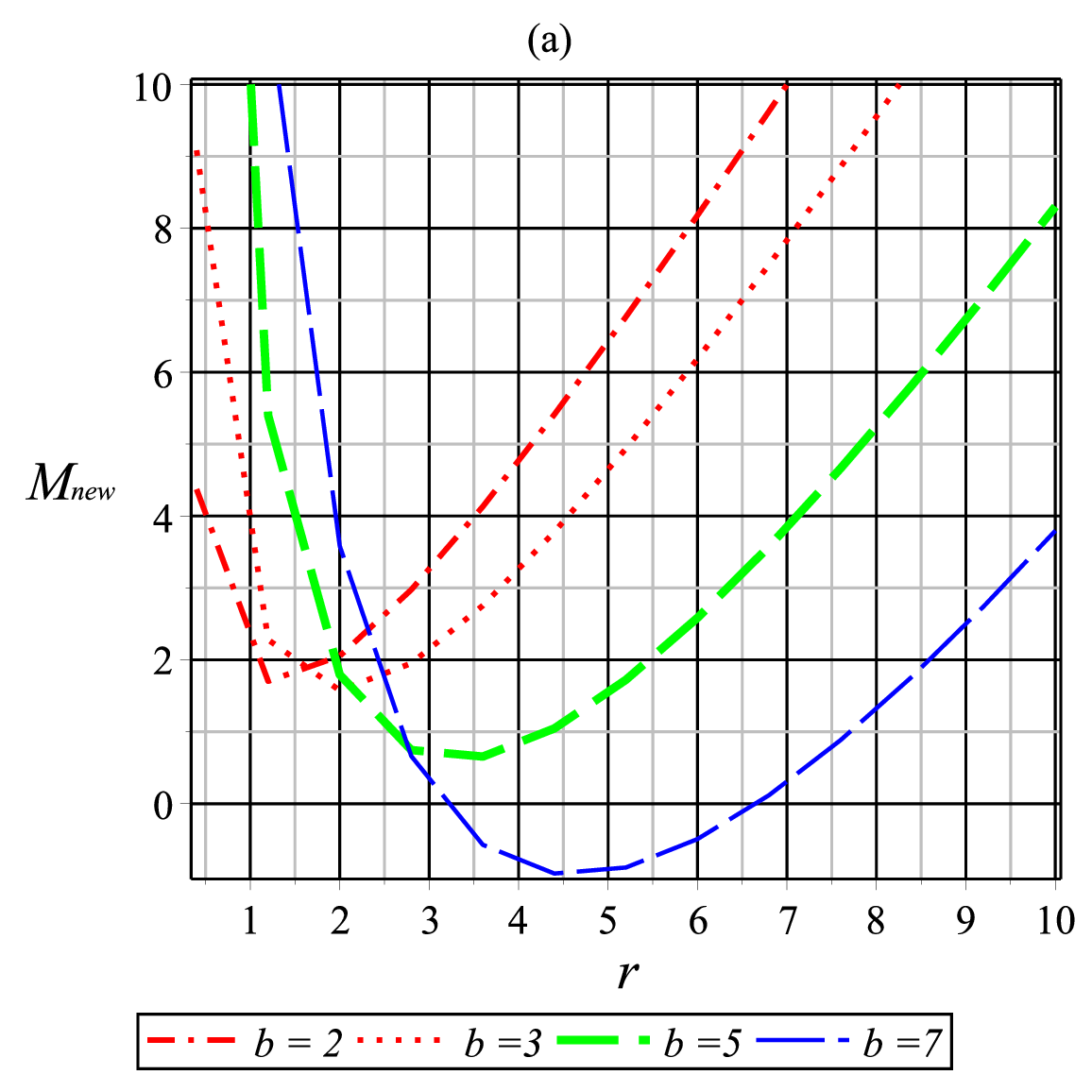}\includegraphics[width=60 mm]{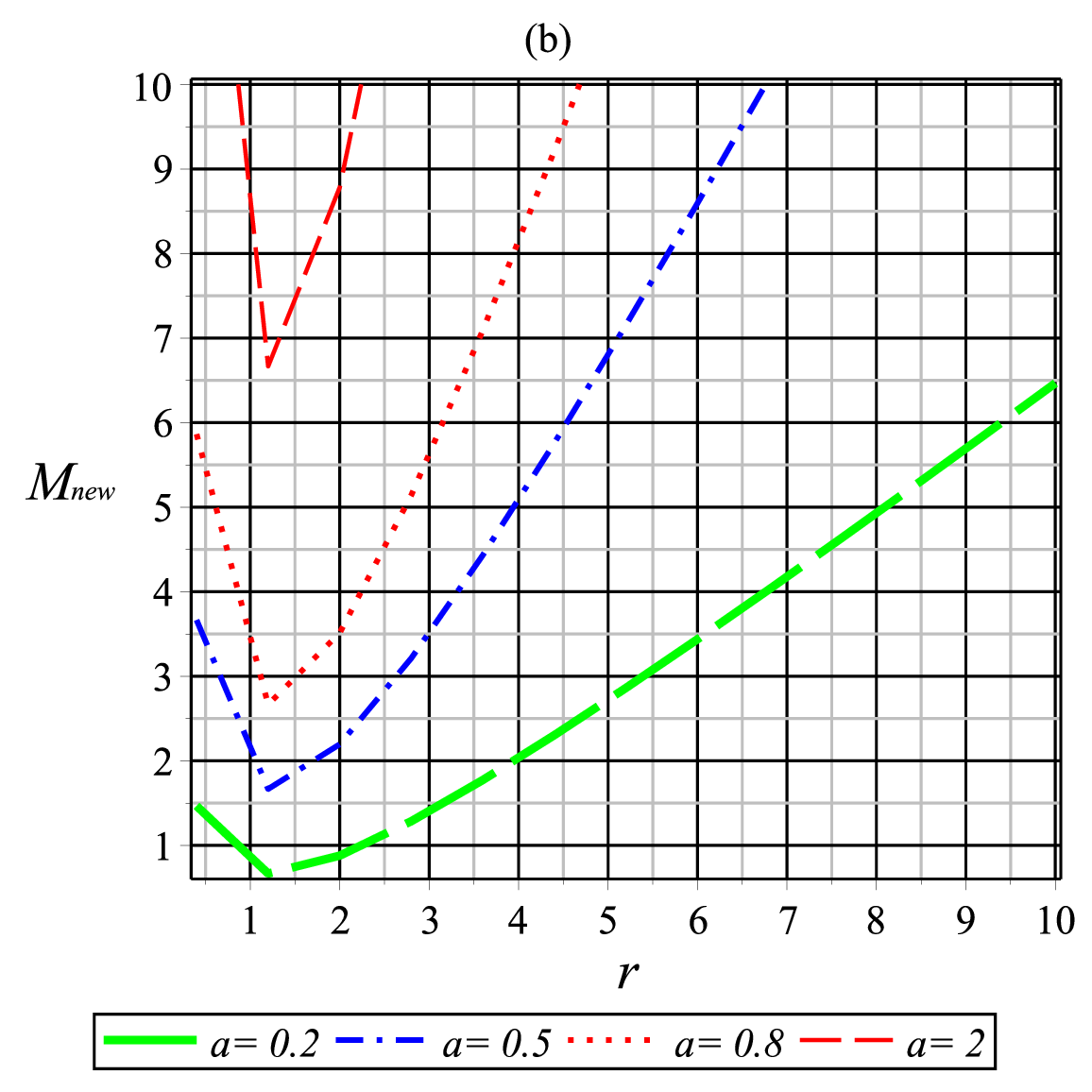}
\end{array}$
\end{center}
\caption{The new physical mass  for, (a)  $a=0.5$, and all possible values of $b$, (b)   $b=1$, and all possible values of $a$.}
\label{fig6}
\end{figure}

We see in  Fig. \ref{fig6}  the behavior of new physical mass in terms of horizon radius.                                                                                                                                                                                                                             To compare Fig. \ref{fig6} and Fig. \ref{fig2}, one can see that the behavior of the physical mass and new physical mass is the same.

\subsection{Critical Points and Stability }
\begin{figure}
\begin{center}$
\begin{array}{cccc}
\includegraphics[width=60 mm]{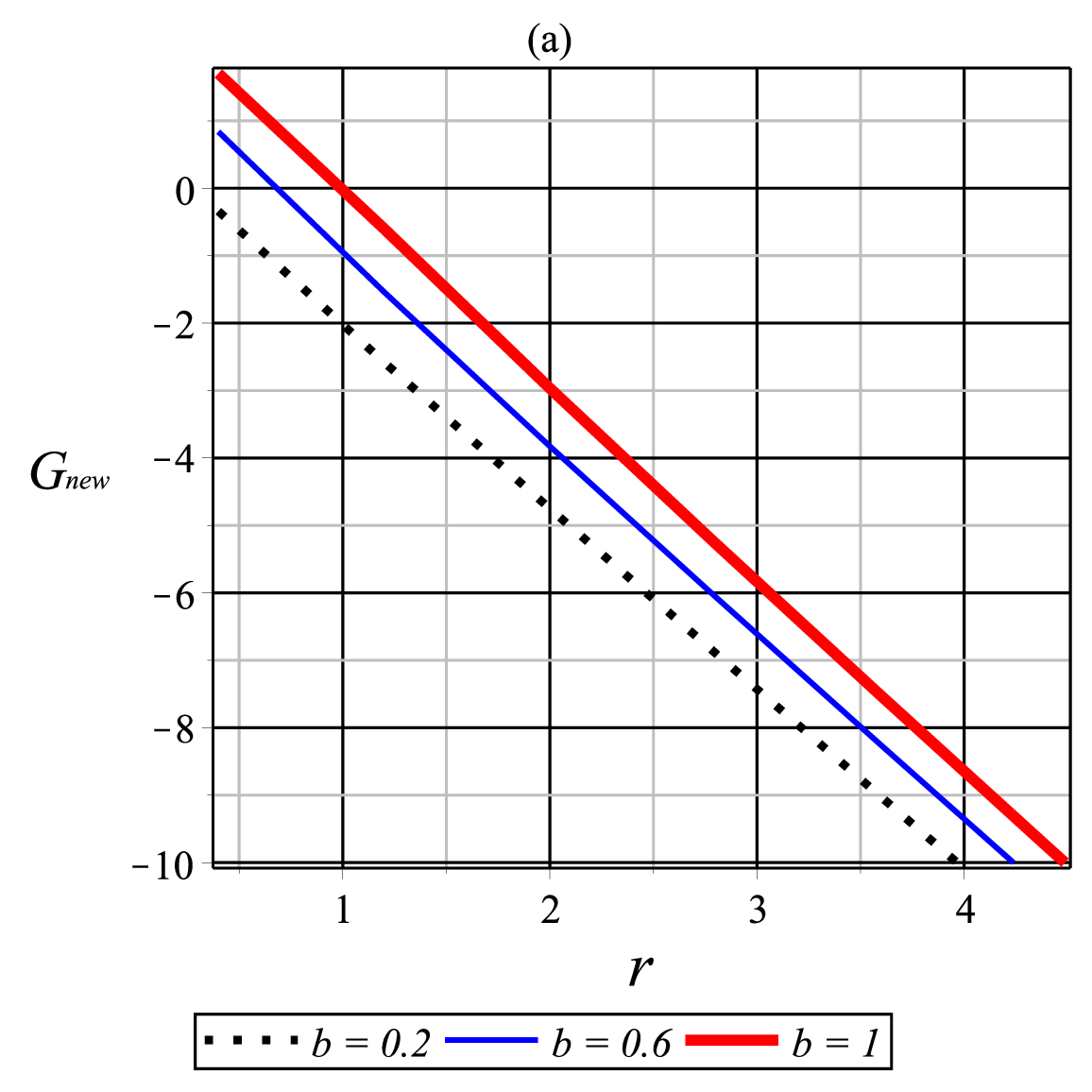}\includegraphics[width=60 mm]{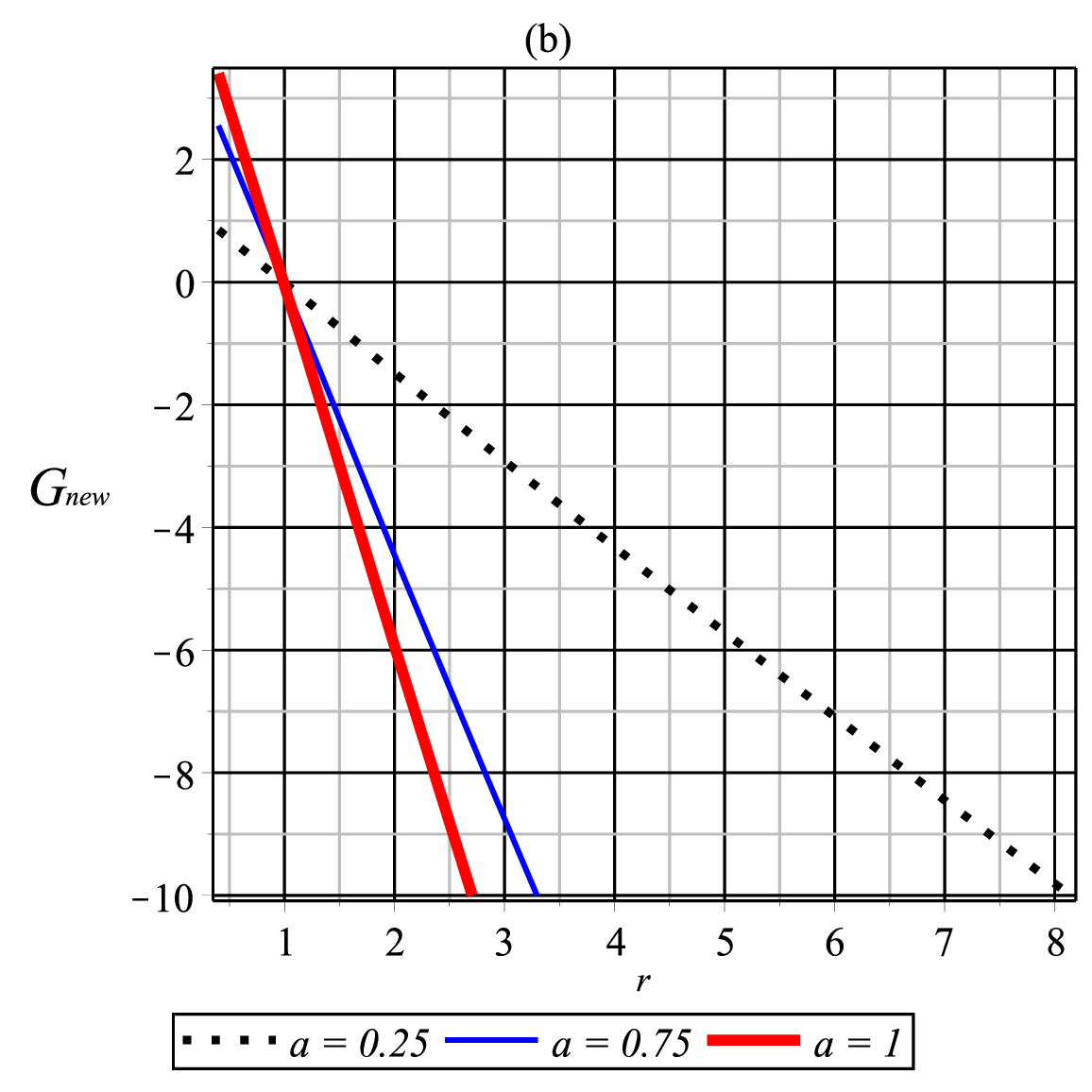}
\end{array}$
\end{center}
\caption{The new Gibbs free energy  (a) all  possible values of $b$,  (b)  all possible values of $a$.}
\label{fig7}
\end{figure}

Now, we want to consider the new stability condition for the corresponding system. In that case,  we need to calculate the new Gibbs free energy  which is given by,
\begin{equation}\label{44}
G_{new} = \frac{23}{12} \pi a b - \frac{5}{3}\pi a r_{+} -   \frac{\pi a b }{2 } \ln {r_{+}} -  \frac{\pi a b^{2}}{4 r_{+}}.
\end{equation}
The graphical analysis of the new Gibbs free energy for modified Horndeski  black hole can be seen in Fig. \ref{fig7}.  We observe in Figs. \ref{fig7}, for the positive  values of $ r_{+} $ the Gibbs free energy has global stability.  Also, to compare Fig. \ref{fig7} and Fig. \ref{fig4},
one can see that  there are a critical point for the usual method. But, this point has been deleted  in the new  Gibbs free energy diagram.
We observe also the Gibbs free energy is unstable.

Also,  the new specific heat is,
\begin{equation}\label{45}
C_{new~P}= \left( \frac{2 S }{3 b} \right)  \frac{4 S - 3 b \sqrt{\pi} S^{\frac{1}{2}}  - 3 \pi  b^{2} }{\sqrt{\pi} S^{\frac{1}{2}} + 2 \pi b }.
\end{equation}
Here, we study two types of phase transition as type one and type two. The phase transition of type one occurs when the specific heat vanishes. It means that $ C_{new}= 0$, which yields to the following equation,
\begin{equation}\label{46}
 4 S - 3 b \sqrt{\pi} S^{\frac{1}{2}}  - 3 \pi  b^{2} =  16 r^{2} _{+} - 6 b r_{+} - 3 b^{2} = 0,
\end{equation}
it gives us the following solution,
\begin{equation}\label{47}
r_{\pm, NC} = \frac{3  \pm \sqrt{57}}{16} b.
\end{equation}
So, the black hole is stable when  horizon radius is $  \frac{3  + \sqrt{57}}{16} b$  and it is unstable when horizon radius is $  \frac{3  - \sqrt{57}}{16} b$.
As we said, the phase transition of type two is associated with divergence points of the specific heat. It means that $\frac{\partial^{2} M}{\partial S^{2}} = 0$. Therefore,
\begin{equation}\label{48}
r_{+} = b
\end{equation}


\begin{figure}
\begin{center}$
\begin{array}{cccc}
\includegraphics[width=55 mm]{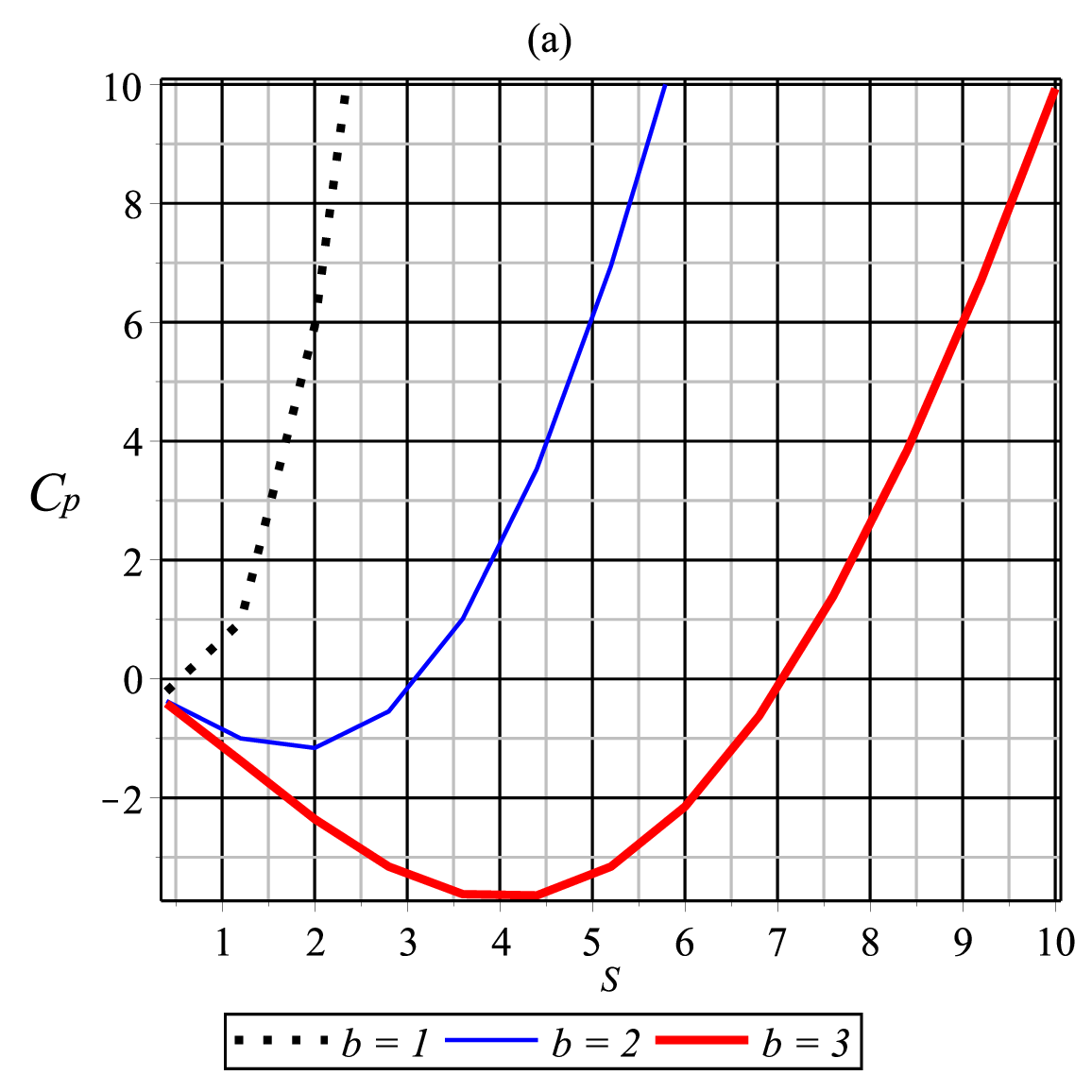}\includegraphics[width=55 mm]{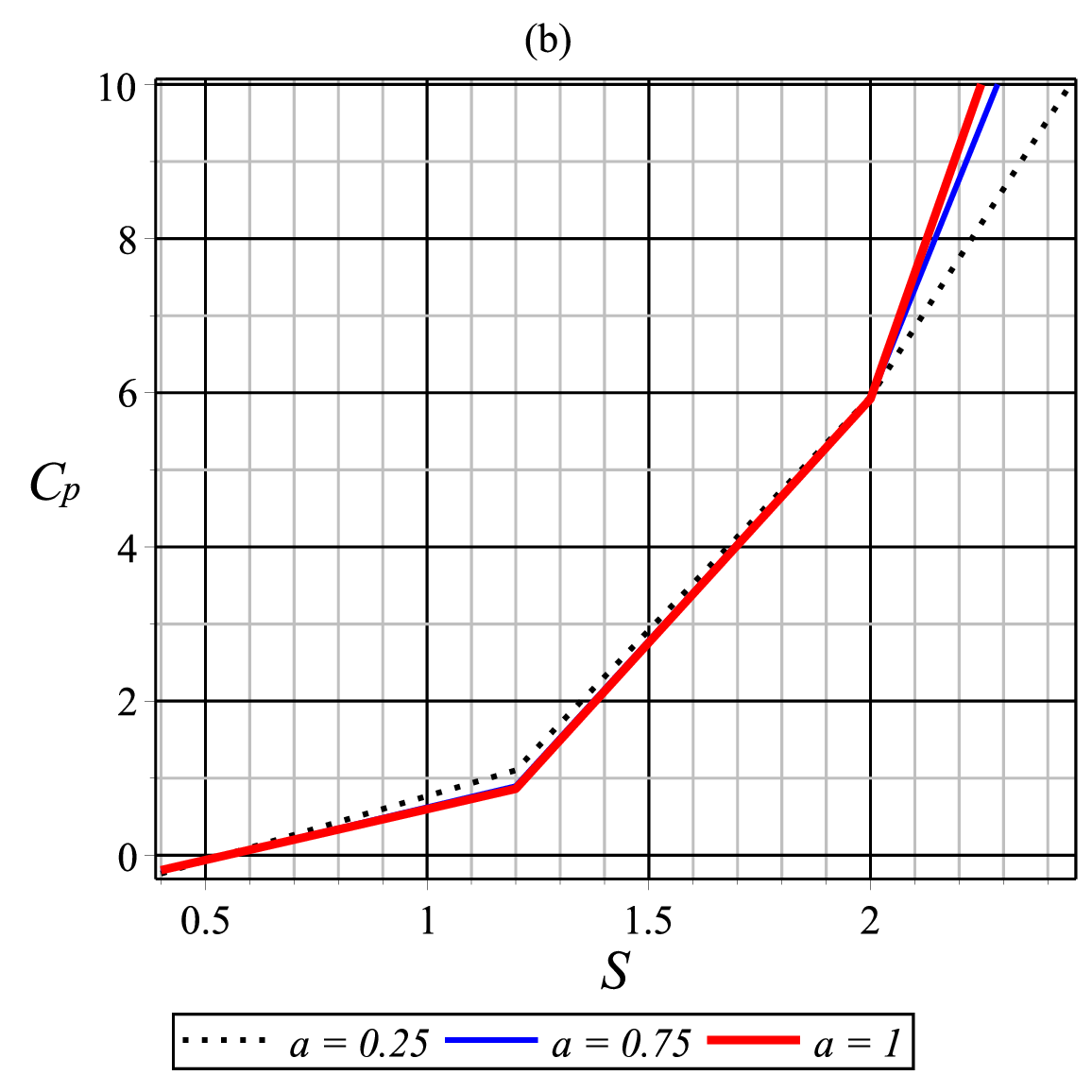}\includegraphics[width=55 mm]{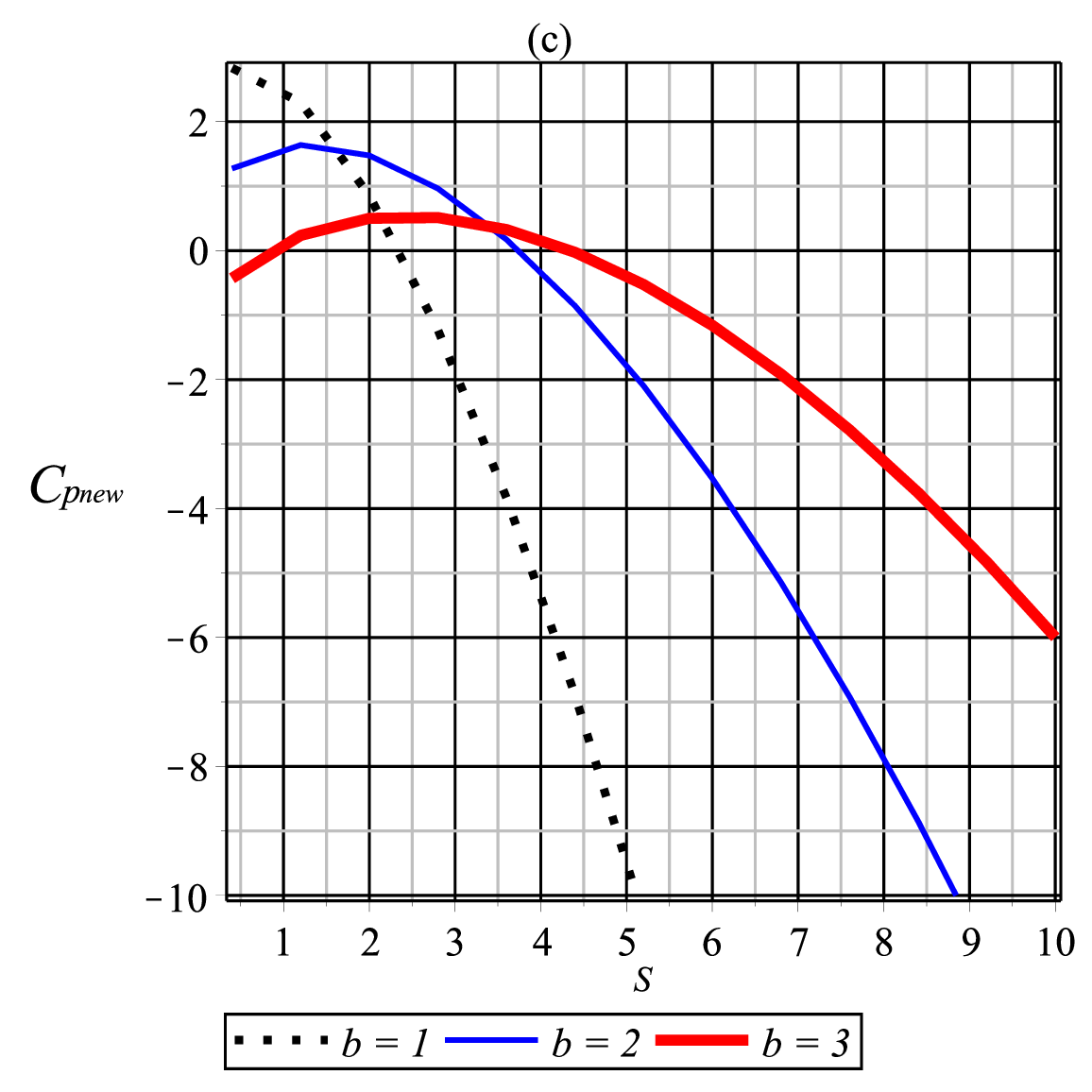}
\end{array}$
\end{center}
\caption{The heat capacity in terms of $ S $ ; (a) all possible  values of $b$ ; (b) all possible values of $a$ ; (c) The new heat capacity in terms of $ S$ for all  possible values of $b$.}
\label{fig8}
\end{figure}

We plot in Fig. \ref{fig8} the behavior of the heat capacity  and show two types of phase transition.
We can see in Fig. \ref{fig8} (a), the heat capacity is decreases when the parameter $ b $ is increases. So that, the unstable state is increases.  We can see that the phase transition occur for different values of $ b $.
Also, In this black hole will have phase transition of type two.\\
As we observe the heat capacity  has stable state for the all values of $ a $ (see Fig. \ref{fig8} (b)).
For a small radius of the black hole, the different  values of parameter $ a $ do not more affect the heat capacity. But, with increasing $ a $ the heat capacity is increases when the horizon radius is large.\\
The graphical analysis of the new heat capacity for modified Horndeski  black hole can be seen in Fig. \ref{fig8} (c)
Hence, we observe that  the new heat capacity in contrast usual case is unstable.\\

It is evident that obtained relation for the pressure (\ref{40}) is different from the usual equation of state (\ref{25}). The maximum is where the system reaches the phase transition.
In order to obtain the maximum of this relation, we use the mathematical nature of extremum points,
\begin{equation}\label{49}
\left( \frac{\partial P_{new}}{\partial r_{+} }\right)_{r_{+} = r_{NC}} = -\frac{2 a}{4 r ^{3}_{NC}} + \frac{3 a b}{4 r ^{4}_{NC} } = 0
\end{equation}
The solution of equation is satisfied by the condition (\ref{46}), which is given by,
\begin{equation}\label{50}
2 r_{NC} = 3b.
\end{equation}
It is exactly  same as the obtained result in usual method. To  replacing $ r_{NC} $ in equations (\ref{40}) - (\ref{43}) we will have,
\begin{eqnarray}\label{51}
T_{NC} &=&  \frac{8 a}{ 27 b},\nonumber\\
P_{NC} &=& \frac{a}{54 b^{2}},
\end{eqnarray}
and also,
\begin{equation}
M_{NC} = \frac{1}{2} \left (\frac{ 19}{6} -  \ln {\frac{3 b}{2}} \right ) \pi a b.
\end{equation}
Finally, the new critically  Gibbs free energy is,
\begin{equation}
G_{NC} = -\frac{1}{4} \left( 3 + 2 \ln {\frac{3 b}{2}} \right) \pi a b.
\end{equation}

\section{Conclusion}
In this paper, we  studied the black hole Horndeski gravity in $ P - V $ critical point.  We checked  the usual  Horndeski black hole  have not $ P - V $ critical behavior.  The above mentioned phase transition lead us,  to applied some anstaz  which is  give us  the modified new metric function $ f(r)$.
This new metric  help us to investigate the critical point for the $P_{c},$ $ T_{c}, $ and $V_{c},$ and the obtained results are complectly the same as van der Waals fluid.  So, we  shown that the  modified Horndeski black hole is satisfied by the equation of state liquid- gas phase transition.
Our main goal of this paper is finding the $ P - V $ criticality  behavior of the modified Horndeski black hole.   First, we obtained  the modified Horndeski black hole by new definition. We showed that there exists a critical point for $ f(r_{+})$ in the corresponding black hole which decreases as long as $M$ (mass of the black hole) increases (see Fig. \ref{fig1}).\\
In order to understand the details of the behavior of the  physical mass, we draw the
corresponding diagrams in Fig. \ref{fig2}.
Here, we said  that  the values of physical mass  decreases and increases before and after the critical horizon radius respectively. We showed that there exists a minimum point for physical  mass  which decreases as long as the coefficients of van der Waals fluid  increases. We  fix the coefficients  $a $ and $  P $ which corresponds to  a van der Waals fluid.
As the parameter $ b $ increases, the critical point of the physical mass is shifted to the right. In this case, we can call $ b $ as a correction quantity. The temperature graph shows that changes in the critical point occur where the horizon radius  is small. We showed the relationship between different coefficients with temperature.  By increasing   coefficients in van der Waals, there is no change in the temperature graph, but the temperature at the critical point decreases.
Also, the temperature at the critical point is divergent when the coefficient $a$ increases.
The Hawking temperature diagram   related to the modified Horndeski black hole was exist in a critical point which  decreases and increases before and after the critical horizon radius respectively.\\
As we know, the  first - order phase transition is where  the temperature is zero.  We obtained  the  first- order phase transition in the points $ r_{+} = 0.41$ and $  r_{+} = 1.71 $ .
One way to check the stability of the system is to calculation of the Gibbs free energy. We found that the coefficients of the van der Waals increase  the stability regions of the black hole  when the radius is very small. But,  the free energy of Gibbs is completely in the state of global stability when the radius is large.
Another way to study  the stability of the system is to calculate the specific heat.
Using specific heat, we found that coefficient $ b $ reduced stable regions of the black hole, but increasing the coefficient $ a $  does not have much effect on the heat capacity. The specific heat reach to stable state when horizon radius is large.\\
Finally, we applied a new method to study phase transition points in this black hole.
Although in the usual method, the phase transition study originates from the temperature related to  the state equation, but the new method is based on the temperature gradient slope of entropy. This new method is a complete method for studying the critical behavior of a thermodynamic system.
The results of the new method are similar to other methods, but they provide more information on the critical behavior of thermodynamic systems that we cannot extract  through other methods. Also, the analytical interpretation of possible phase transition points lead us to arrange some nonphysical range of horizon radius for the corresponding black hole.
Another advantage of this method is that it discusses all thermodynamic quantities.
It will be interesting to analyze the effect of the thermal fluctuations to modified black holes on this new method. And compared its results with the usual method \cite{28,29,30,31}.

\end{document}